\documentclass[twocolumn]{aastex63} 

\usepackage{amsmath}
\usepackage{hyperref}
\usepackage{amssymb}
\usepackage{empheq}
\usepackage{mathrsfs}
\usepackage{epstopdf}
\usepackage{enumitem}
\usepackage{graphicx}
\AppendGraphicsExtensions{.pdf}				
\usepackage{amssymb}

\shorttitle{Stellar Obliquity Distributions for Hot and Warm Jupiters}
\shortauthors{Morgan et al.}
\graphicspath{{./}{figures/}}
\begin{document}


\title{Signs of Similar Stellar Obliquity Distributions for Hot and Warm Jupiters Orbiting Cool Stars}

\correspondingauthor{Marvin Morgan}
\email{marv08@utexas.edu }

\author[0000-0003-4022-6234]{Marvin Morgan}
\affiliation{Department of Astronomy, The University of Texas at Austin, Austin, TX 78712, USA}

\author[0000-0003-2649-2288]{Brendan P. Bowler}
\affiliation{Department of Astronomy, The University of Texas at Austin, Austin, TX 78712, USA}

\author[0000-0001-6532-6755]{Quang H. Tran}
\affiliation{Department of Astronomy, The University of Texas at Austin, Austin, TX 78712, USA}

\author[0000-0003-0967-2893]{Erik Petigura}
\affiliation{Department of Physics and Astronomy, University of California Los Angeles, Los Angeles, CA 90095, USA}

\author[0000-0001-5909-4433]{Vighnesh Nagpal}
\affiliation{Department of Astronomy, University of California, Berkeley, CA 94720, USA}

\author[0000-0002-3199-2888]{Sarah Blunt}
\affiliation{Department of Astronomy, California Institute of Technology, Pasadena, CA, USA}

\begin{abstract}
Transiting giant planets provide a natural opportunity to examine stellar obliquities, which offer clues about the origin and dynamical histories of close-in planets. Hot Jupiters orbiting Sun-like stars show a tendency for obliquity alignment, which suggests that obliquities are rarely excited or that tidal realignment is common.  However, the stellar obliquity distribution is less clear for giant planets at wider separations where realignment mechanisms are not expected to operate. In this work, we uniformly derive line-of-sight inclinations for 47~cool stars ($T_\mathrm{eff}$ $<$ 6200~K) harboring transiting hot and warm giant planets by combining rotation periods, stellar radii, and $v \sin i$ measurements. Among the systems that show signs of spin-orbit misalignment in our sample, three are identified as being misaligned here for the first time. Of particular interest are Kepler-1654, one of the longest-period (1047 d;~2.0~AU) giant planets in a misaligned system, and Kepler-30, a multi-planet misaligned system. By comparing the reconstructed underlying inclination distributions, we find that the inferred minimum misalignment distributions of hot Jupiters spanning $a/R_{*}$~=~3--20 ($\approx$~0.01--0.1~AU) and warm Jupiters spanning $a/R_{*}$~=~20--400 ($\approx$~0.1--1.9~AU) are in good agreement. With 90$\%$ confidence, at least 24$^{+9}_{-7}\%$ of warm Jupiters and 14$^{+7}_{-5}\%$ of hot Jupiters around cool stars are misaligned by at least 10$^\circ$. Most stars harboring warm Jupiters are therefore consistent with spin-orbit alignment. The similarity of hot and warm Jupiter misalignment rates suggests that either the occasional misalignments are primordial and originate in misaligned disks, or the same underlying processes that create misaligned hot Jupiters also lead to misaligned warm Jupiters.

\end{abstract}

\section{Introduction}\label{sec:intro}
Our Solar System contains two gas giants and two ice giants on coplanar and near-circular orbits at distances beyond 5 AU. The discovery of 51 Peg b, a 0.5 $M_\mathrm{Jup}$ planet on a 4.5-day orbit (\citealt{1995Natur.378..355M}), and several other early planetary detections (\citealt{1996ApJ...464L.147M}; \citealt{1997ApJ...483..457C}; \citealt{2000ApJ...529L..45C}), planted the first seeds of a blossoming new field of astronomy. These discoveries overhauled the established understanding of planetary formation, migration, and orbital architectures. It is now clear from planet period and eccentricity distributions that substantial orbital evolution is common, and perhaps even ubiquitous among giant planets (\citealt{Winn2015}).

The giant planets in our Solar System reside outside of the ``water ice line," the location in a protoplanetary disk where water condenses into solid ice. The solar ice line is currently located in the asteroid belt at $\sim$2–3 AU (\citealt{2008ApJ...673..502K}). Beyond this region, rapid planetesimal and core growth is facilitated, which results in more efficient assembly of giant planets (\citealt{2006ApJ...640.1115L}). Long-baseline radial velocity (RV) surveys have found that giant planets are prevalent at orbital distances of 1--10 AU compared to orbits interior or exterior of this range (\citealt{2019ApJ...874...81F}; \citealt{2021ApJS..255...14F}). Direct imaging surveys have also found similar results that show giant planets are less abundant at wider orbital distances (\citealt{Bowler2016}; \citealt{Baron2019}; \citealt{Nielsen2019}).  Although the occurrence rate of giant planets appears to peak beyond the location of the water ice line, there remains a significant population of gas giants at closer separations. The origin and evolution of these Jovian  planets within $\sim$ 2 AU of Sun-like stars has proven to be challenging to observationally constrain.

 Several mechanisms have been proposed to explain the presence of giant planets interior to the water ice line. Early interactions with the protoplanetary disk can result in inward migration (\citealt{1980ApJ...241..425G};  \citealt{1996Natur.380..606L};  \citealt{2012ARA&A..50..211K}). Some gas giants at close separations may have formed in situ if favorable conditions are met (\citealt{2016ApJ...829..114B}). Kozai-Lidov (KL) oscillations with an outer companion represent another viable mechanism for giant planets to migrate inward when coupled with high-eccentricity tidal damping (\citealt{1962AJ.....67..591K}; \citealt{1962P&SS....9..719L}; \citealt{2003ApJ...589..605W}; \citealt{2016ARA&A..54..441N}). When eccentricities are excited and periastron distances shrink during these oscillations, tidal friction can dissipate orbital energy and circularize the planet's orbit, breaking the KL cycles and freezing the planet’s orbital parameters (\citealt{2001ApJ...562.1012E}; \citealt{2007ApJ...669.1298F}; \citealt{2007ApJ...670..820W}).  Planet-planet scattering can also trigger high eccentricity migration through a secular or chaotic exchange of angular momentum between planets (\citealt{1996Sci...274..954R}; \citealt{Chatterjee2008};  \citealt{2012ApJ...751..119B};  \citealt{DawsonJohnson2018}). 

 High-eccentricity tidal migration driven by KL oscillations or planet-planet scattering is a leading pathway to produce hot Jupiters (\citealt{Triaud2010}; \citealt{Albrecht2012}). However, this process can only occur if the planet passes close enough to its host star to gravitationally interact with the stellar envelope. Warm Jupiters, situated beyond $\sim 0.1$ AU, are too far from their host star to raise dissipative tides. \citet{Dawson2015} and \citet{Jackson2023} place an upper limit on KL oscillations as a viable migration mechanism for most hot and warm Jupiters due to a lack of observed highly-eccentric proto-hot Jupiters by Kepler. Proto-hot Jupiters orbiting bright, more metal-rich, nearby stars observed by TESS, such as TOI-3362 b, may not be as uncommon (\citealt{Dong2021}).

The relative alignment of the stellar rotation axis and the planetary orbital plane can provide complementary insight into inward giant planet migration processes of warm Jupiters. A variety of mechanisms can misalign these rotational and orbital angular momentum vectors during or after the era of giant planet formation. Torques induced from binary companions and primordial disk structures can cause the misalignment of the stellar spin axis. \citet{2022PASP..134h2001A} found that primordial misalignments might be produced by stellar flybys that occur during the epoch of planet formation. In this scenario, stellar companions can induce torques on protoplanetary disks, which can give rise to spin-orbit misalignments with any planets that eventually form (\citealt{Batygin2012}; \citealt{2013ApJ...778..169B}; \citealt{Spalding2014}). \citet{Epstein-Martin2022} found that broken and misaligned disks are capable of torquing the spin axis of their host star. Interactions between stellar magnetic fields and circumstellar disks may also be able to generate a broad distribution of spin-orbit angles (\citealt{Lai2011}).

To date, most obliquity measurements have been constrained from the Rossiter–McLaughlin (RM) effect which measures the sky-projected spin–orbit angle between a star’s equatorial plane and a transiting planet's orbital plane, $\lambda$ (\citealt{1924ApJ....60...15R}; \citealt{1924ApJ....60...22M}). HD 209458 b was the first exoplanet for which this phenomenon was reported (\citealt{2000A&A...359L..13Q}), laying the foundation for over 100 additional measurements (\citealt{2022PASP..134h2001A}). Most RM measurements have been obtained for hot Jupiters as they have frequent transits, large RM-induced RV amplitudes, and favorable geometric transit probabilities. RM measurements of hot Jupiters have revealed that misalignments are common around hot stars but less frequent around cool stars below the Kraft break ($T_\mathrm{eff}$ $<$ 6200 K), which might be a result of tidal realignment (\citealt{2009ApJ...696.1230F}; \citealt{Triaud2010}; \citealt{2010ApJ...719..602S}; \citealt{Albrecht2012}).

In contrast, few RM measurements have been obtained for long-period transiting warm Jupiters due to a combination of their infrequent transits, small geometric transit probabilities, and long transit durations.\footnote[1]{HIP 41378 d, a Neptune-sized transiting exoplanet with an orbital period of 278 days, is the longest-period planet of any size with an RM measurement (\citealt{Grouffal2022}).} The longest-period giant planet for which the RM effect has been measured is HD 80606 b, a transiting warm Jupiter with an orbital period of 111.44 days and $a/R_{*}$ = 94.64 (\citealt{Pont2009}; \citealt{2022PASP..134h2001A}). \footnote[2]{TOI-1859 b ($a/R_{*}$ = 53.7) is the second-longest-period giant planet for which the RM effect was measured (\citealt{Dong2023}).} Moving to larger orbital distances provides unique constraints on migration channels as it removes the possibility for tidal circularization, realignment, and synchronization and thus probes alternative migration and misalignment mechanisms. 

 \citet{2022AJ....164..104R} found evidence that in single-star systems, warm Jupiters may be preferentially more aligned than hot Jupiters. They attribute this to differences in the formation and migration of hot and warm Jupiters. However, \citet{Albrecht2012} found hints of an opposite trend, where hot Jupiters are mostly consistent with alignment while warm Jupiters in their sample have significant misalignments. More recent studies have used  starspot-induced amplitudes to identify a correlation between increased misalignment with orbital separation moving outward to 50-day orbital periods (\citealt{Mazeh2015}; \citealt{LiWinn2016}). 

If RM measurements are not available, a lower limit on the true obliquity, $\psi$, can be determined using the inclination of the host star in combination with the inclination of a transiting planet  (\citealt{2014ApJ...783....9H}; \citealt{MortonWinn2014}; \citealt{MasudaWinn2020}). In this work, we investigate the minimum inferred stellar obliquities of stars hosting transiting giant planets beyond 0.1 AU. Minimum misalignment distributions are inferred from homogeneous and self-consistent measurements of $i_{*}$, the line-of-sight stellar spin inclination. Together with knowledge of the transiting planet's orbital geometry, this provides information about $\psi$. Here, we explore a simple question: are warm Jupiter host stars misaligned at similar rates  as hot Jupiter host stars? Establishing whether these two populations are similar or distinct can provide valuable clues about giant planet inward migration timescales and mechanisms.

This paper is organized as follows. In Section \ref{sec:Target_Selection} we discuss our target selection criteria and describe how we construct our hot and warm Jupiter samples. In Section \ref{sec:Alignment_Analysis} we describe our process of measuring individual and population-level stellar inclination distributions. In Section \ref{sec:Discussion} we present our results and discuss interpretations of the hot and warm Jupiter stellar obliquity distributions. Next, we describe individual misaligned systems in Section \ref{sec:Individual_Systems}. Finally, we summarize our conclusions in Section \ref{sec:Conclusion}.

\begin{figure}
\begin{center}
{\includegraphics[width=\linewidth]{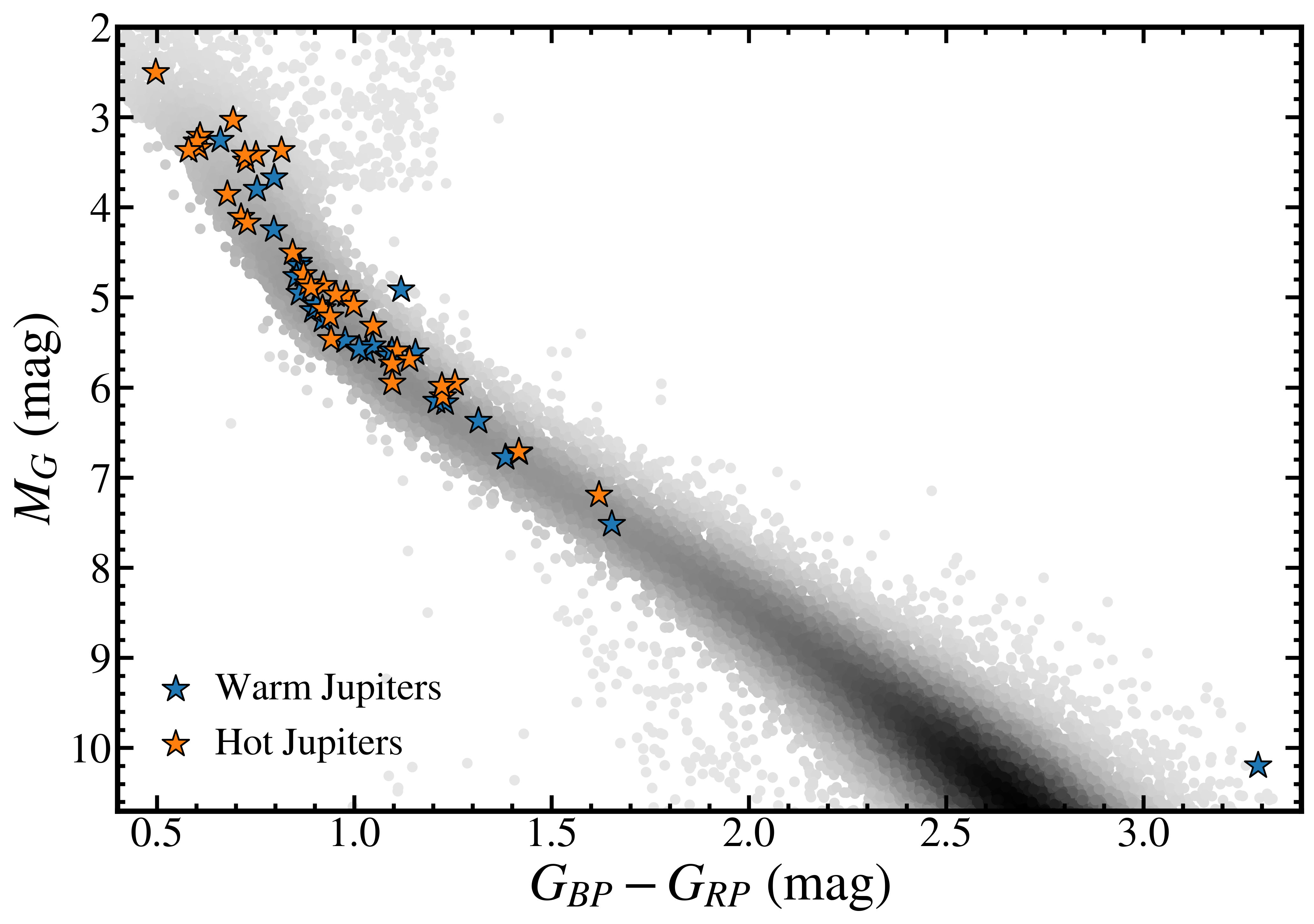}}
\caption{Planet-hosting stars in our sample plotted on a Gaia $M_{G}$ vs $G_{BP}$ - $G_{RP}$ color-magnitude diagram. Host stars are color coded based on whether they harbor a hot Jupiter (orange) or a warm Jupiter (blue). Most systems in our analysis are main sequence FGK dwarfs. The star with the reddest color is TOI-1227, a young (11 $\pm$ 2 Myr), low-mass (0.17 $M_{\odot}$) member of the Lower Centaurus Crux OB association.}
\label{fig:Gaia_CMD}
\end{center}
\end{figure}

\section{Target Selection and Rotation Periods}\label{sec:Target_Selection}

Our sample of warm Jupiters originates from the NASA Exoplanet Archive (\citealt{2013PASP..125..989A}), as of August 2022. We selected transiting planets with either a measured minimum mass of $m_p \sin i$ = 0.3--13 $M_\mathrm{Jup}$, or a radius $R_p > 8 R_{\oplus}$, ensuring that low-mass brown dwarfs and sub-Jovian sized planets are excluded. We then isolated systems with scaled orbital distances of $a/R_{*}$ $>$ 20 to probe giant planets with semi-major axes $\gtrsim$ 0.1 AU. For a Sun-like star with a field age of several Gyr, a separation of $a/R_{*}$ $>$ 20 corresponds to a realignment timescale greater than the system age and a circularization timescale of $>$ $1$ Gyr (\citealt{RasioTides1996}; \citealt{Lubow1997}; \citealt{MurrayDermott2000}; \citealt{Husnoo2012}; \citealt{Spalding2022}). This cut in $a/R_{*}$ reflects where warm Jupiters are expected to be largely undisturbed by tidal forces, in comparison to hot Jupiters which may have experienced more dynamically violent tidal migration and spin-orbit realignment (\citealt{2022AJ....164..104R}).\footnote[3]{Note that a threshold of $a/R_{*}$ = 20 can include giant planets with orbital periods under 10 days depending on the mass and radius of the host star.} This resulted in 110 transiting warm Jupiters orbiting 104 host stars. These systems represent our initial sample to measure stellar inclinations and minimum stellar obliquities, which is only possible for a subset of stars with rotation periods and projected rotational velocities.  

When available, we analyze \emph{TESS}, \emph{Kepler}, and \emph{K2} light curves of warm Jupiter host stars in our sample to uniformly determine rotation periods.  To avoid confusion between pulsations from $\gamma$ Dor variables and rotation periods from spot-driven modulations, only stars with spectral types of F5 or later are considered here. We use the \texttt{lightkurve} \citep{Lightkurve2018} software package to search for and download all available 30-minute Pre-search Data Conditioning Simple Aperture Photometry \citep[PDCSAP;][]{Jenkins2010}, 30-minute K2 extracted light curves \citep[K2SFF;][]{Howell2014, Vanderburg2014}, and 2-minute cadence \emph{TESS} Science Processing Operations Center (SPOC) PDCSAP \citep{Smith2012, Stumpe2012, Stumpe2014, Jenkins2016} light curves for each target from the Mikulski Archive for Space Telescopes (MAST) data archive.\footnote[4]{\href{http://archive.stsci.edu/kepler/data_search/search.php}
{http://archive.stsci.edu/kepler/data\_search/search.php}, \href{https://archive.stsci.edu/k2/data_search/search.php}{https://archive.stsci.edu/k2/data\_search/search.php} \href{https://archive.stsci.edu/missions-and-data/tess}{https://archive.stsci.edu/missions-and-data/tess/}} Individual \textit{Kepler} quarters, \textit{K}2 campaigns, and \textit{TESS} sectors are then normalized and stitched together to create the final light curves (see the Notes section of Table \ref{tab:host_stars} for details).\footnote[5]{Some of the data presented in this paper were obtained from MAST at the Space Telescope Science Institute and can be accessed via \dataset[10.17909/xk84-vr57]{https://doi.org/10.17909/xk84-vr57}.} Finally, flares, transits, and other outliers are removed by running a high-pass Savitzky-Golay filter \citep{Savitzky1964} through the light curve and selecting data points lying within three sigma of the photometric average.

For each normalized light curve, we compute a Generalized Lomb-Scargle periodogram \citep[GLS;][]{Zechmeister2009} over the frequency range $0.0005-100.0$ d$^{-1}$ ($0.01-2000$ days) to search for any rotational modulation. Periods and uncertainties were measured by fitting a Gaussian to the highest periodogram peak. In the case where there is a large envelope resulting from fringe patterns in the GLS periodogram, the Gaussian was fit to the total curve in order to reflect that spread. Targets with a single strong peak, whose phase-folded light curves showed clear periodicity, and amplitudes of both the periodogram and rotational modulation were large are considered to have reliable period measurements. 

There are two host stars, K2-290 and WASP 84, with rotation period measurements adopted from \citet{Hjorth2019} and \citet{Anderson2014A}  respectively, which satisfy our initial selection cuts but did not show clear periodic brightness variations in their  \emph{TESS} or \emph{Kepler} light curves. We adopt the published rotation periods for these two systems.

We then retrieve $v \sin i$  and stellar radii measurements from the literature. Published projected rotational velocities are compiled, and a weighted mean of available measurements is adopted following the procedure in \citet{Bowler2023}. The spectra of K2-281, K2-77, Kepler-486, Kepler-52, and Kepler-1654 were obtained with the HIRES spectrometer (\citealt{Vogt1994}) on the 10-m Keck-I Telescope between 2012-2018. The spectra were observed as part of several reconnaissance efforts to characterize \emph{Kepler} and \emph{K2} planet-hosting stars by the California Planet Search (\citealt{Howard2010}) described in \citet{Petigura2017} and \citet{Sinukoff2018}. Spectral S/N ranged from 22-45 per reduced pixel on blaze at 5500 $\AA$. We used \texttt{Specmatch-Syn} code described in \citet{Petigura2015} to determine $v \sin i$. At this S/N, \texttt{SpecMatch-Syn} returns $v \sin i$ measurements with uncertainties of 1 km s$^{-1}$ when $v \sin i$ is larger than 2 km s$^{-1}$. When $v \sin i$ is lower, the results are upper limits with $v \sin i$ $<$ 2 km s$^{-1}$.

All hot Jupiters in our sample, including the parameters for their host stars, are obtained from \citet{2022PASP..134h2001A} and have measured minimum masses of $m_p \sin i$ = 0.3--13 $M_\mathrm{Jup}$ and $a/R_{*}$ $<$ 20. We further filter the sample based on binary architecture. Close binaries with P-type circumbinary planets (planets orbiting around more than one host star) are removed, as migration channels may differ in these dynamically complex systems. Altogether, this yielded samples of 36 transiting hot Jupiters and 24 transiting warm Jupiters with measured rotation periods, $v \sin i$ values, and radius estimates for their host stars.

\begin{figure}
\begin{center}
{\includegraphics[width=\linewidth]{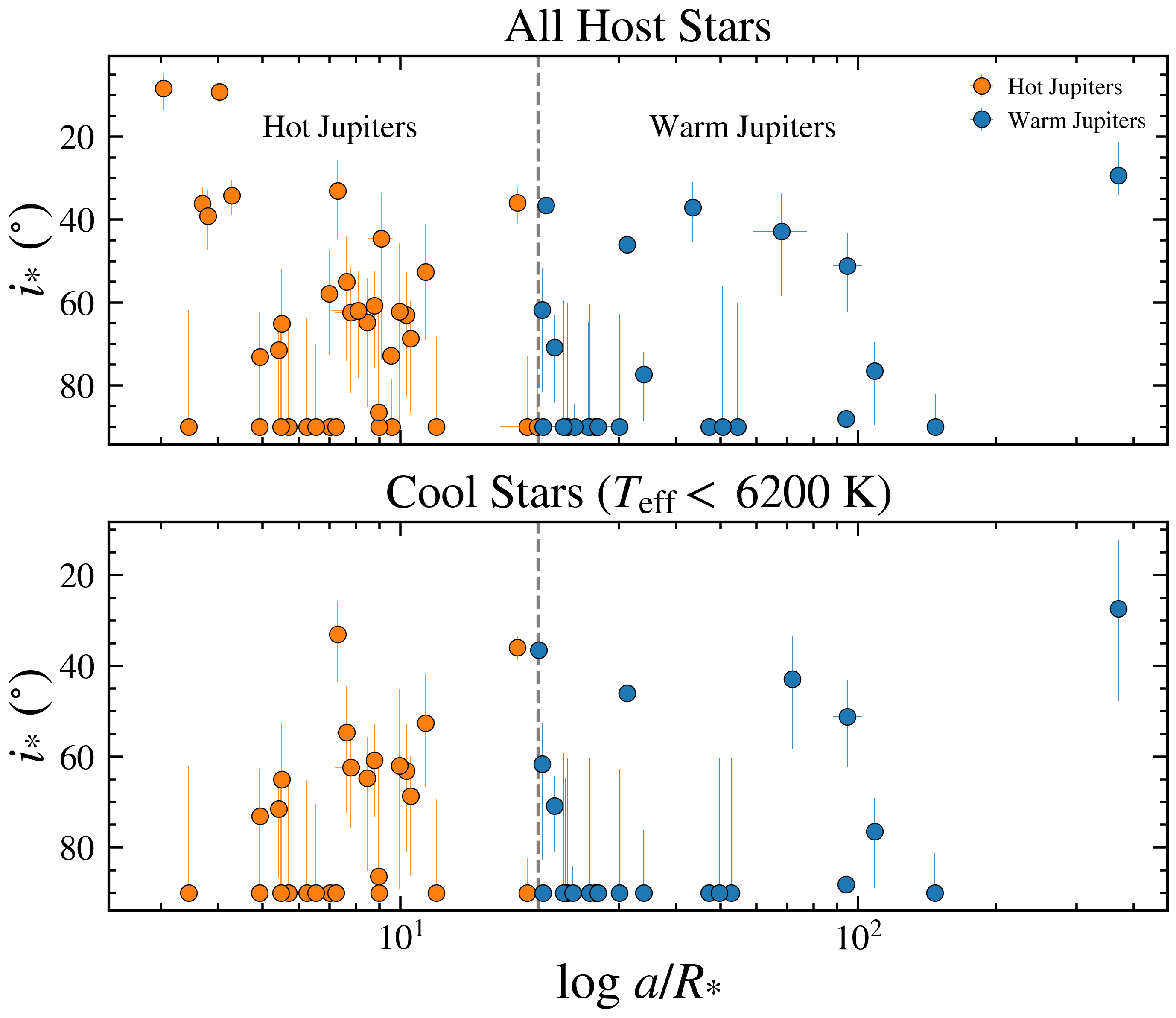}}
\caption{Measured $i_{*}$ values for hot (orange) and warm (blue) Jupiters plotted as a function of $a/R_{*}$. The dashed line is the separation between hot and warm Jupiters for $a/R_{*}$ of 20 around a Sun-like star. The $y$-axis spans 90$^\circ$ to 0$^\circ$ to emphasize aligned systems at $i_*$ = 90$\degr$. The top panel shows all of the 36 hot Jupiters and 24 warm Jupiters compiled for our analysis. In the bottom panel, we have isolated the sample to 25 hot Jupiter and 22 warm Jupiter host stars below the Kraft break.}
\label{fig:i_vs_a_comparison_vals}
\end{center}
\end{figure}

To generate a consistent comparison between the hot and warm Jupiter sample, we have made an additional cut to focus on cool stars with $T_\mathrm{eff}$ $<$ 6200 K (see Section \ref{sec:Alignment_Analysis}). This effective temperature corresponds to the Kraft break, a gradual transition between stars that experience Sun-like spin-down and stars that experience little to no angular momentum loss (\citealt{2022ApJ...930....7A}). \citet{1967ApJ...150..551K} discovered that hot stars with thin outer convective zones cannot support magnetized winds while cool stars with $T_\mathrm{eff}$ $\lesssim$ 6200 K experience substantial angular momentum loss due to the presence of large convection zones and strong winds. A \emph{Gaia} color-magnitude diagram of our full sample of host stars can be seen in Figure \ref{fig:Gaia_CMD}. The final number of hot and warm Jupiters orbiting cool host stars with $P_{\rm{rot}}$, $v \sin i$, and $R_{*}$ constraints is 25 and 22, respectively, as shown in Figure \ref{fig:i_vs_a_comparison_vals}. 

\section{Results}\label{sec:Alignment_Analysis}

\subsection{Stellar Inclinations}

Our approach to infer stellar inclinations follows the Bayesian framework from \citet{MasudaWinn2020}. They considered the relationship between stellar equatorial velocity, 2$\pi$$R_*$/$P_\mathrm{rot}$, and 
the projected rotational velocity, $v \sin i$, while properly accounting for the correlation between these parameters. \citet{Bowler2023} derived analytical expressions for the stellar inclination posterior
$P(i_*)$ assuming uniform priors on $v \sin i$, $R_*$, and $P_\mathrm{rot}$;
an isotropic ($\sin i_*$) prior on stellar inclination; 
and a moderately precise constraint on the stellar rotation period ($\sigma_{P_\mathrm{rot}}/P_\mathrm{rot} \lesssim$20\%):

\begin{equation}{\label{eqn:i}}
\resizebox{\hsize}{!}{$p(i_*\mid P_\mathrm{rot}, R_*, v\sin i_*) \propto \sin i_* \times \frac{e^{- \frac{\left(v \sin i_* - \frac{2\pi R_*}{P_\mathrm{rot}}\sin i_* \right)^2}{2\left(\sigma_{v\sin i_*}^2 + \sigma_{v_\mathrm{eq}}^2 \sin^2 i_* \right)}} }{\sqrt{\sigma_{v\sin i_*}^2 + \sigma_{v_\mathrm{eq}}^2 \sin^2 i_*}},$}
\end{equation}

where

\begin{equation}{\label{eqn:veqerr}}
\sigma_{v_\mathrm{eq}} =  \frac{2\pi R_*}{P_\mathrm{rot}} \sqrt{\Big(\frac{\sigma_{R_*}}{R_*}\Big)^2 + \Big(\frac{\sigma_{P_\mathrm{rot}}}{P_\mathrm{rot}}\Big)^2 }.
\end{equation}

\noindent Here $\sigma_{P_\mathrm{rot}}$, $\sigma_{v \sin i}$, and $\sigma_{R_*}$ are the uncertainties on the rotation period, projected rotational velocity,
and stellar radius.  

Differential rotation will cause starspots located at mid-latitudes to travel faster than the equatorial velocity.  This can bias rotation periods inferred from light curves (e.g., \citealt{ReinholdGizon2015}).  To account for these potential systematic errors, we inflate the nominal rotation period uncertainty  from our light curve periodogram analysis following \citet{Bowler2023}.  Assuming a Sun-like pole-to-equator absolute shear of 0.07 rad day$^{-1}$, this typically increases the rotation period uncertainty by a factor of $\approx$3 (with a range of 1--70). 

Line-of-sight stellar inclination posteriors are determined in this fashion for hot and warm Jupiter host stars in our sample using new and compiled $v \sin i$ values, rotation periods, and radius estimates (Table \ref{tab:new_vsini}; Table \ref{tab:full_table}).  
In one instance for Kepler-1654, the host star is a slow rotator and the $v \sin i$ value is only constrained to $<$2~km s$^{-1}$. In this case we use Equation A17 from \citet{Bowler2023}, which accounts for rotational broadening as an upper limit:


\begin{equation}{\label{eqn:i_ul}}
\resizebox{\hsize}{!}{$p(i_* \mid P_\mathrm{rot}, R_*, v\sin i_*) \propto \sin i_* \times \Big( \mathrm{erf}\Big(\frac{l - \frac{2\pi R_*}{P_\mathrm{rot}} \sin i_*}{\sqrt{2} \, \sigma_{v_\mathrm{eq}} \sin i_*}  \Big) + \mathrm{erf}\Big(\frac{\sqrt{2} \pi R_*}{\sigma_{v_\mathrm{eq}} P_\mathrm{rot} }  \Big)   \Big),$}
\end{equation}

\noindent where $l$ is the upper limit on the projected rotational velocity and erf is the error function.

\begin{figure*}
  \vskip -0.1in
  \hskip 0.2in
  \resizebox{0.9\textwidth}{!}{\includegraphics{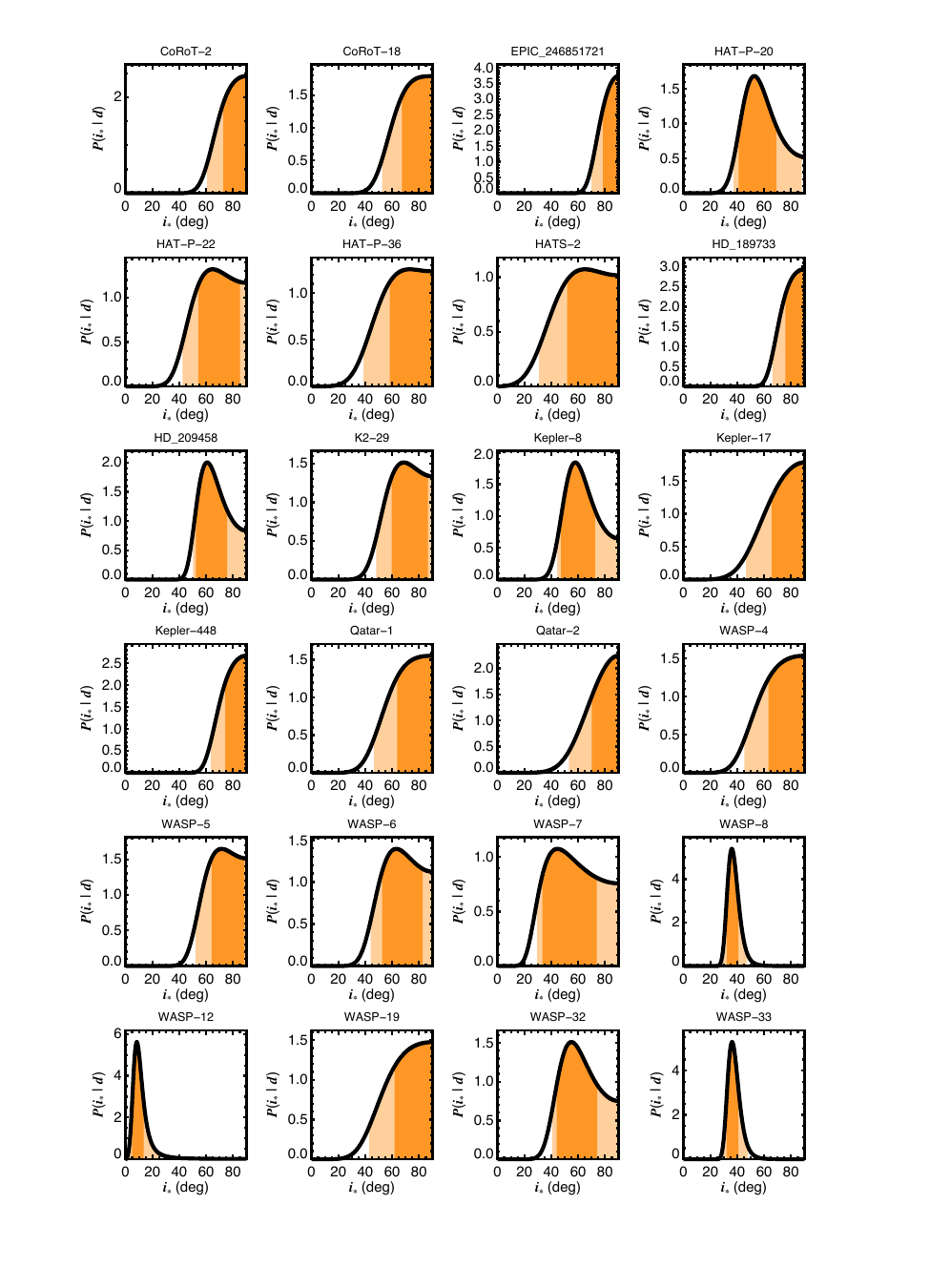}}
  \vskip -0.4in
  \caption{Posterior distributions of stellar inclination for stars hosting hot Jupiters in our sample. Shaded regions show the $68\%$ and $95\%$ credible intervals.\label{fig:obliquities_pg1}} 
\end{figure*}

\begin{figure*}
  \vskip -0.1 in
  \hskip 0.2 in
  \resizebox{0.9\textwidth}{!}{\includegraphics{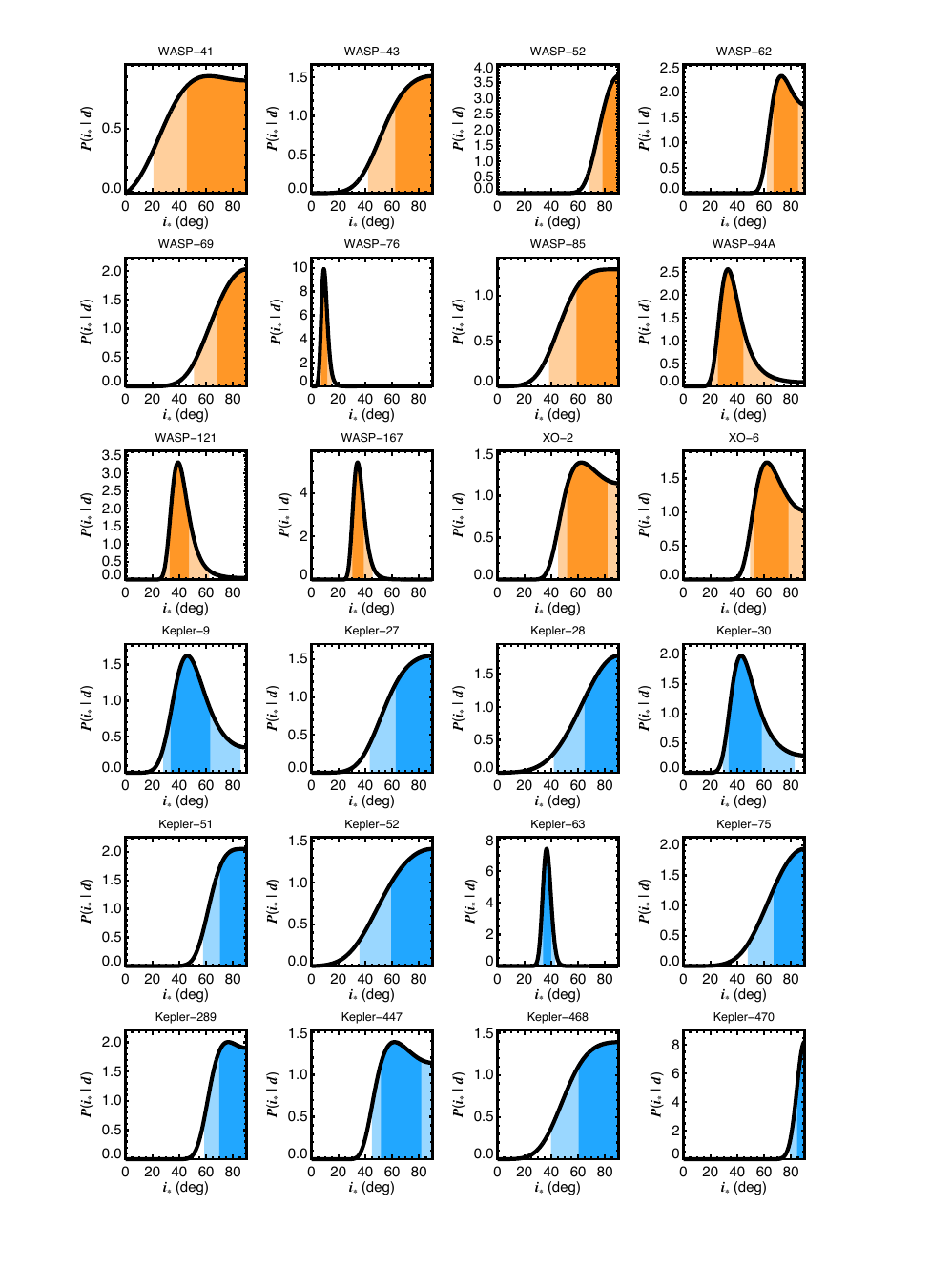}}
  \vskip -0.4 in
  \caption{Posterior distributions of stellar inclination for stars hosting hot (orange) and warm (blue) Jupiters in our sample.  Shaded regions show the 68\% and 95\% credible intervals. \label{fig:obliquities_pg2} } 
 \end{figure*}

\begin{figure*}
  \vskip -0.1 in
  \hskip 0.2 in
  \resizebox{6.40in}{!}{\includegraphics{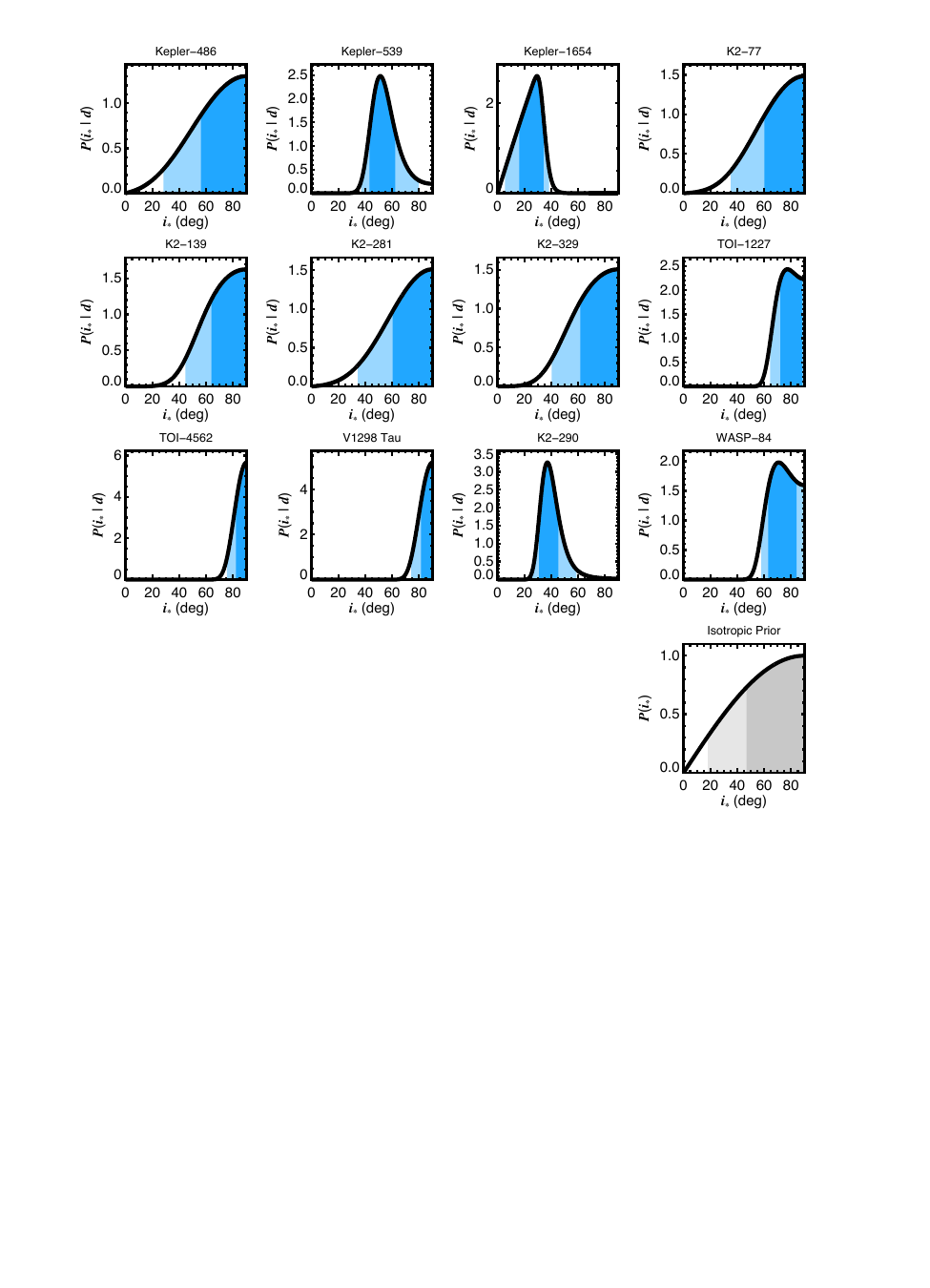}}
  \vskip -3.2 in
  \caption{Posterior distributions of stellar inclination for stars hosting  warm Jupiters in our sample.  Shaded regions
  show the 68\% and 95\% credible intervals.  The $\sin i_*$ isotropic prior distribution is shown for comparison in the bottom right panel. \label{fig:obliquities_pg3} } 
 \end{figure*}

Results for all 61 host stars are shown in Figures~\ref{fig:obliquities_pg1}--\ref{fig:obliquities_pg3} and summary statistics for each distribution can be found in Table \ref{tab:full_table}.  There are a wide variety of constraints; in some cases there is only a small departure from the $\sin i_*$ isotropic prior, while in other cases inclinations are constrained to within a few degrees.  Overall these results are in good agreement with previous stellar inclination measurements. For instance, \citet{2022PASP..134h2001A} found $i_{*}$ = 90$^{+0}_{-11}\degr$ for CoRoT-2 and  73$^{+12}_{-6}\degr$ for WASP-62 while we derive $i_{*}$ = 90$^{+0.1}_{-8}\degr$ and 73$^{+11}_{-6}\degr$, respectively. 

It is immediately evident from these posterior distributions which systems host misaligned planets.  Any distribution that departs from $i_{*}$ = 90$\degr$ implies a minimum misalignment by at least that difference because transiting planets have orbital inclinations of $\approx$90$\degr$. We note, however, that there could be misaligned systems in this sample that do not have host stars with inclinations that depart from 90$\degr$ because the true obliquity angle also depends on the polar position angle of the star and the longitude of ascending node of the planet's orbit.
Many systems stand out as being significantly misaligned.  Some of these are previously known such as HAT-P-20 (\citealt{Esposito2017}) and Kepler-63 (\citealt{Sanchis-Ojeda2013}) while several are newly identified in this work, including Kepler-539 b, with an orbital period of 125 days ($a/R_{*}$ = 94.61; \citealt{Mancini2016}) and Kepler-1654 b, which orbits at 1047 days ($a/R_{*}$ = 370.3; \citealt{2018AJ....155..158B}).

For this study we have adopted the following classification for aligned and misaligned systems. Host stars that have a maximum a posteriori probability (MAP) value $>$10$^\circ$ with 90$\%$ confidence are classified as being \emph{misaligned}. Hosts that have a MAP value $>$10$^\circ$ with 80$\%$ confidence are \emph{likely misaligned}. Following this framework, 7 out of 25 hot Jupiters around cool stars are either misaligned or likely misaligned.  For the warm Jupiter sample, 6 out of 22 stars are misaligned or likely misaligned. We find with $90\%$ confidence that the probability for any particular warm Jupiter host star to be misaligned by at least $i_{*}$ = 10$^\circ$ is $24^{+9}_{-7}\%$. We also find with $90\%$ confidence the probability for any particular hot Jupiter host star to be misaligned by at least $i_{*}$ = 10$^\circ$ is $14^{+7}_{-5}\%$. These results are summarized in Table  \ref{tab:misalignment_percentiles}. 

\subsection{Hierarchical Bayesian Analysis}\label{sec:Bayesian_Modeling}

Hierarchical Bayesian modeling (HBM) offers a natural framework to simultaneously infer parameters of individual systems and hyperparameters governing the underlying behavior of a population. In this study we follow the sampling approach outlined in \citet{Hogg2010}. We infer the underlying distribution of $i_{*}$ values for our sample of hot and warm Jupiter hosts assuming a flexible population-level parametric model.

 The hot and warm Jupiter samples are separately analyzed using \texttt{ePop!}, an open source Python package for fitting population-level distributions to sets of individual system distributions (\citealt{Nagpal2022}). Our adopted underlying model is the Beta Distribution, a set of continuous probability distributions constrained on the interval [0,1] with two free parameters $\alpha$ and $\beta$: 
 \begin{equation}
    \label{beta_equation}
      B(x)=\frac{\Gamma(\alpha+\beta)}{\Gamma(\alpha) \Gamma(\beta)}e^{\alpha-1}(1-e)^{\beta-1} .
\end{equation}

For this analysis, each minimum obliquity distribution is first re-mapped onto a new variable $\theta'$ = $\theta/90 \degr$ so as to span a range of 0--1 rather than 0--90$\degr$. In the framework of HBM, $\alpha$ and $\beta$ become hyperparameters whose posterior distributions are constrained using the affine-invariant Markov chain Monte Carlo sampler \texttt{emcee} (\citealt{Foreman-Mackey2013}).\footnote[5]{We also re-parameterized the Beta distribution following \citet{Dong2023b}, where $\alpha$ = $\mu$*$\kappa$ and $\beta$ = (1-$\mu$)*$\kappa$, and reproduce similar results with a uniform hyperprior spanning 0 to 1 on $\mu$ and a log-normal hyperprior on $\kappa$ centered on 0 with a standard deviation of 3. Results for hot and warm Jupiter hosts are consistent, reinforcing the similarity of the two populations.} To test the impact of our choice of hyperpriors on the $\alpha$ and $\beta$ posteriors, we carry out two fits using different hyperpriors on each parameter: a Truncated Gaussian with $\mu$ = 0.69, $\sigma$ = 1.0
 \begin{equation}p(x) = \frac{1}{\sigma}\frac{\frac{1}{\sqrt{2\pi}} e^{-\frac{1}{2}\left(\frac{x-\mu}{\sigma}\right)^{2}}}{1-\frac{1}{2} \left(1+\mathrm{erf} \left(\frac{-\mu}{\sigma \sqrt{2}}\right)\right)} , 
\end{equation}
 and a log-uniform distribution ranging from 0.01 to 100 (\citealt{Nagpal2022})
\begin{equation} p(x) = \frac{1}{x}\ .
\end{equation}

\begin{figure}
\begin{center}
{\includegraphics[width=\linewidth]{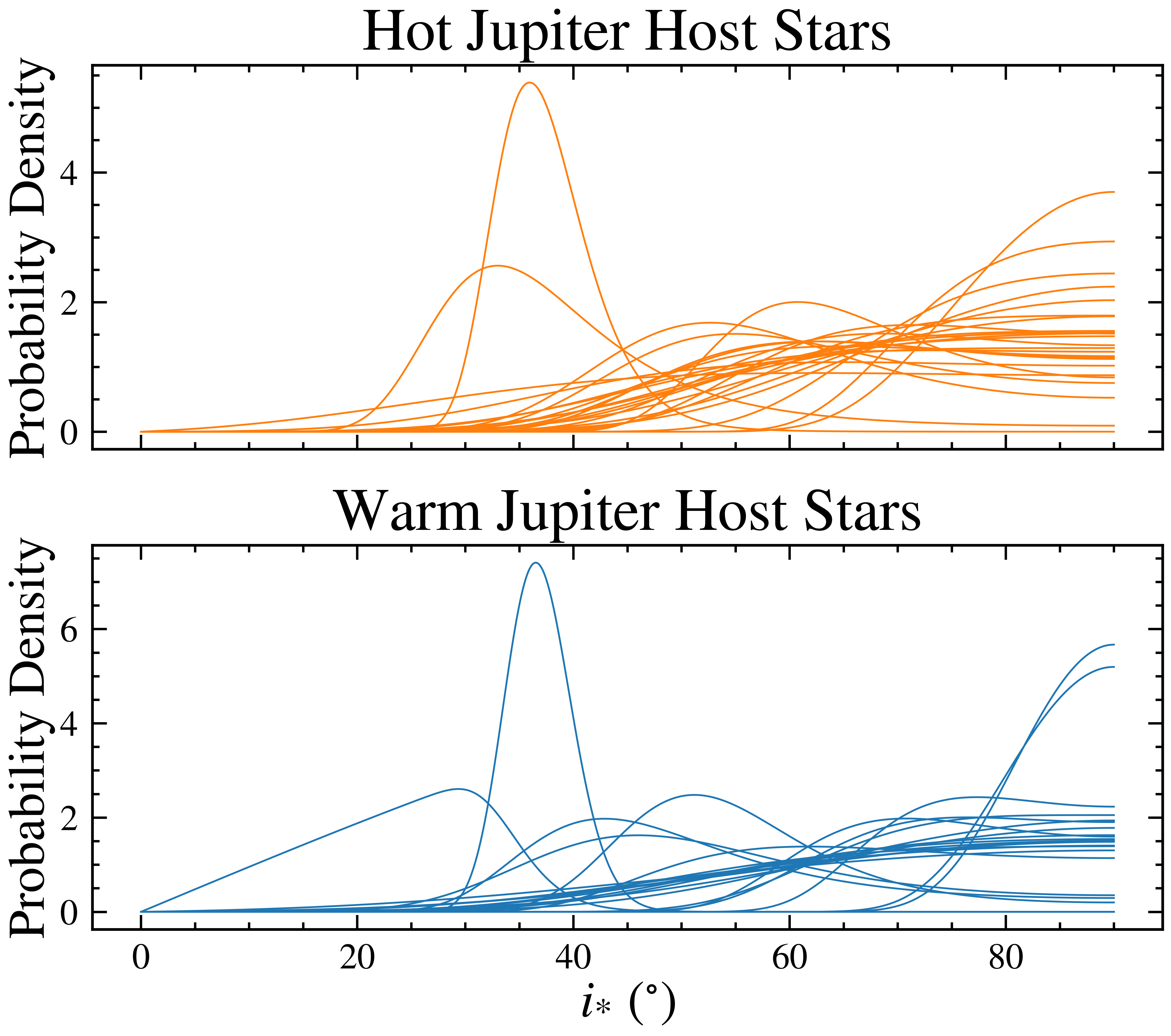}}
\caption{$i_*$ posterior distributions for the 25 hot Jupiter ($a/R_{*}$ $<$ 20; top panel) and 22 warm Jupiter ($a/R_{*}$ $>$ 20; bottom panel) host stars below the Kraft break. These $i_*$ posterior distributions are then used as individual likelihood functions in \texttt{ePop!}.}
\label{fig:PDF_vals1}
\end{center}
\end{figure}

A burn-in fraction of 50$\%$ is adopted and 50 walkers are run for 5 $\times$ 10$^4$ steps. The best-fit posterior values are reported in Table \ref{tab:Model_Parameters}. An overview of stellar inclination posteriors for the hot and warm Jupiter samples can be found in Figure \ref{fig:PDF_vals1}.

\startlongtable
\begin{deluxetable}{lcl}
\renewcommand\arraystretch{0.9}
\tabletypesize{\footnotesize}
\setlength{\tabcolsep} {.1cm} 
\tablewidth{0pt}
\tablecolumns{3}
\tablecaption{Adopted Projected Rotational Velocities.\label{tab:new_vsini}}
\tablehead{
     & \colhead{$v \sin i$} &    \\
       \colhead{Name}  & \colhead{(km s$^{-1}$)}  &  \colhead{Reference} 
        }   
\startdata
Kepler-9     &    1.1  $\pm$  1.0  &   \cite{2017AJ....154..107P}  \\
Kepler-9     &    2.2  $\pm$  0.5  &   \citet{2012Natur.486..375B}    \\
Kepler-9     &    2.3 &    \cite{2016ApJS..225...32B}     \\
\hline
\textbf{Kepler-9}  &   \textbf{2.0  $\pm$  0.4}   &     \textbf{Adopted} \\
\hline
Kepler-51    &   5.4   $\pm$ 0.6   &   \citet{2018ApJS..237...38B}    \\
Kepler-51    &   5.5   $\pm$ 1.0    &  \cite{2017AJ....154..107P}    \\
Kepler-51   &    6.83  &    \citet{2020AJ....160..120J}     \\
\hline
\textbf{Kepler-51}  &   \textbf{5.4  $\pm$  0.5}   &     \textbf{Adopted} \\
\hline
Kepler-289  &    5.5  $\pm$  0.5   &   \citet{2018ApJS..237...38B}   \\
Kepler-289  &    5.8  $\pm$  1.0   &   \citet{2017AJ....154..107P}  \\
Kepler-289   &   5.2   &    \cite{2016ApJS..225...32B}    \\
\hline
\textbf{Kepler-289}  &   \textbf{5.6  $\pm$  0.4}   &     \textbf{Adopted} \\
\hline
Kepler-447   &   7.3  $\pm$  0.5  &    \citet{2018ApJS..237...38B}     \\
Kepler-447   &   6.9  $\pm$  1.0  &    \citet{2017AJ....154..107P}    \\
Kepler-447   &   7.47  &    \citet{2020AJ....160..120J}    \\
\hline
\textbf{Kepler-447}  &   \textbf{7.2  $\pm$  0.4}   &     \textbf{Adopted} \\
\hline
Kepler-539   &   3.5  $\pm$  0.5   &   \citet{2018ApJS..237...38B}       \\
Kepler-539   &   3.0  $\pm$  1.0   &   \citet{2017AJ....154..107P}     \\
Kepler-539   &   2.8  $\pm$  0.5   &   \citet{2012Natur.486..375B}    \\
Kepler-539  &    6.54  &     \citet{2020AJ....160..120J}   \\
\hline
\textbf{Kepler-539}  &   \textbf{3.1  $\pm$  0.3}   &     \textbf{Adopted} \\
\hline
Kepler-1654 &    0.3   &     \citet{2016ApJS..225...32B}    \\
Kepler-1654   &  $< 2.0$  &     \citet{2018AJ....155..158B}   \\
Kepler-1654  & $< 2.0$  & This work \\
\hline
\textbf{Kepler-1654}  &   \textbf{$< 2.0$}   & 
\textbf{Adopted} \\
\hline
V1298 Tau   &     24.10 $\pm$ 1.4   &   \citet{2012ApJ...745..119N}     \\
V1298 Tau   &     24.87 $\pm$ 0.19  &   \citet{2022AJ....163..247J}       \\
\hline
\textbf{V1298 Tau}  &   \textbf{24.8 $\pm$  0.2}   &     \textbf{Adopted} \\
\hline
K2-77      &     4   &     \citet{2017MNRAS.464..850G}   \\
K2-77      &     2.9 $\pm$    1.0  &    This work \\
\hline
\textbf{K2-77}  &   \textbf{2.9 $\pm$    1.0}   &     \textbf{Adopted} \\
\hline
K2-139     &     2.8  $\pm$  0.6  &    \citet{2018MNRAS.475.1765B}     \\
K2-139     &     1.7  &     \citet{2018AJ....155...21P}     \\
\hline
\textbf{K2-139}  &   \textbf{2.8  $\pm$  0.6}   &     \textbf{Adopted} \\
\hline
K2-281      &    3  $\pm$  1  &   This work   \\%
\hline
\textbf{K2-281}  &   \textbf{3  $\pm$  1}   &     \textbf{Adopted} \\
\hline
TOI-4562    &    17  $\pm$  0.5   &   \citet{2022arXiv220810854H}   \\
TOI-4562    &    15.7  $\pm$  0.5   &   \citet{2022arXiv220810854H}   \\
TOI-4562    &    16.5  $\pm$  0.56   &   \citet{Sharma2018}   \\
\hline
\textbf{TOI-4562}  &   \textbf{16.4  $\pm$  0.3}   &     \textbf{Adopted} \\
\hline
TOI-1227   &     16.65  $\pm$  0.24 &   \citet{2022AJ....163..156M}    \\
\hline
\textbf{TOI-1227}  &   \textbf{16.65  $\pm$  0.24}   &     \textbf{Adopted} \\
\hline
Kepler-27    &   2.4    $\pm$  1.0   &  \citet{2017AJ....154..107P}  \\
Kepler-27    &   0.6    $\pm$  5.0   &  \citet{2012MNRAS.421.2342S}   \\
Kepler-27    &   2.76   $\pm$  1.53  &  \citet{2012MNRAS.421.2342S}    \\
Kepler-27    &   3.0    $\pm$  1.0   &  \citet{2022AJ....163..179P}  \\
\hline
\textbf{Kepler-27}  &   \textbf{2.7  $\pm$  0.6}   &     \textbf{Adopted} \\
\hline
Kepler-28   &    3.8   $\pm$   1.0  &   \citet{2017AJ....154..107P}  \\
Kepler-28   &   5.5    &   \citet{2020AJ....160..120J}   \\
\hline
\textbf{Kepler-28}  &   \textbf{3.8  $\pm$  1.0}   &     \textbf{Adopted} \\
\hline
Kepler-30    &   2.3  $\pm$    1.0  &   \citet{2017AJ....154..107P}   \\
Kepler-30    &   1.94  $\pm$   0.22 &   \citet{2012ApJ...750..114F}   \\
Kepler-30    &   2.2   $\pm$   1.0  &  \citet{2022AJ....163..179P}    \\
\hline
\textbf{Kepler-30}  &   \textbf{2.0  $\pm$  0.2}   &     \textbf{Adopted} \\
\hline
Kepler-470   &  20.9  $\pm$  1.0  &   \citet{2022AJ....163..179P}  \\
Kepler-470   &  20.5  &    \citet{2020AJ....160..120J}   \\
\hline
\textbf{Kepler-470}  &   \textbf{20.9  $\pm$  1.0}   &     \textbf{Adopted} \\
\hline
Kepler-486   &  2.2     $\pm$  1.0  &   This work  \\
\hline
\textbf{Kepler-486}  &   \textbf{2.2     $\pm$  1.0}   &     \textbf{Adopted} \\
\hline
Kepler-52   &  3     $\pm$  1  &   This work  \\
\hline
\textbf{Kepler-52}  &   \textbf{3     $\pm$  1}   &     \textbf{Adopted} \\
\hline
Kepler-63  &   5.8  $\pm$   0.5  &   \citet{2018ApJS..237...38B}   \\
Kepler-63  &   4.8  $\pm$   1.0  &   \citet{2017AJ....154..107P}   \\
Kepler-63  &   3.8  $\pm$   0.5  &   \citet{2012Natur.486..375B}  \\
Kepler-63  &   14  $\pm$   3    &    \citet{Frasca2022}    \\
Kepler-63  &   0  $\pm$   3     &   \citet{Frasca2022}   \\
Kepler-63  &   5.43  &   \citet{2019AJ....158..101M}    \\
Kepler-63  &   7.2   &   \citet{2020AJ....160..120J}   \\
Kepler-63  &   5.9  &     \citet{2016ApJS..225...32B}   \\
\hline
\textbf{Kepler-63}  &   \textbf{5.0  $\pm$  0.3}   &     \textbf{Adopted} \\
\hline
Kepler-75   &  3.1  $\pm$   1.0  &   \citet{2017AJ....154..107P}  \\
Kepler-75   &  3.2  $\pm$   1.0  &  \citet{2022AJ....163..179P}   \\
\hline
\textbf{Kepler-75}  &   \textbf{3.2  $\pm$  0.7}   &     \textbf{Adopted} \\
\hline
Kepler-468  &  3.9  $\pm$   1.0 &   \citet{2022AJ....163..179P} \\
\hline
\textbf{Kepler-468}  &   \textbf{3.9  $\pm$   1.0}   &     \textbf{Adopted} \\
\hline
K2-329    &    1.9  $\pm$   0.5  &  \citet{2021AJ....161...82S}  \\
\hline
\textbf{K2-329}  &   \textbf{1.9  $\pm$   0.5}   &  \textbf{Adopted} 
\enddata
\end{deluxetable}

\begin{figure}
  \centering
  
  \begin{minipage}[b]{0.4\textwidth}
    \centering
    \includegraphics[width=\linewidth]{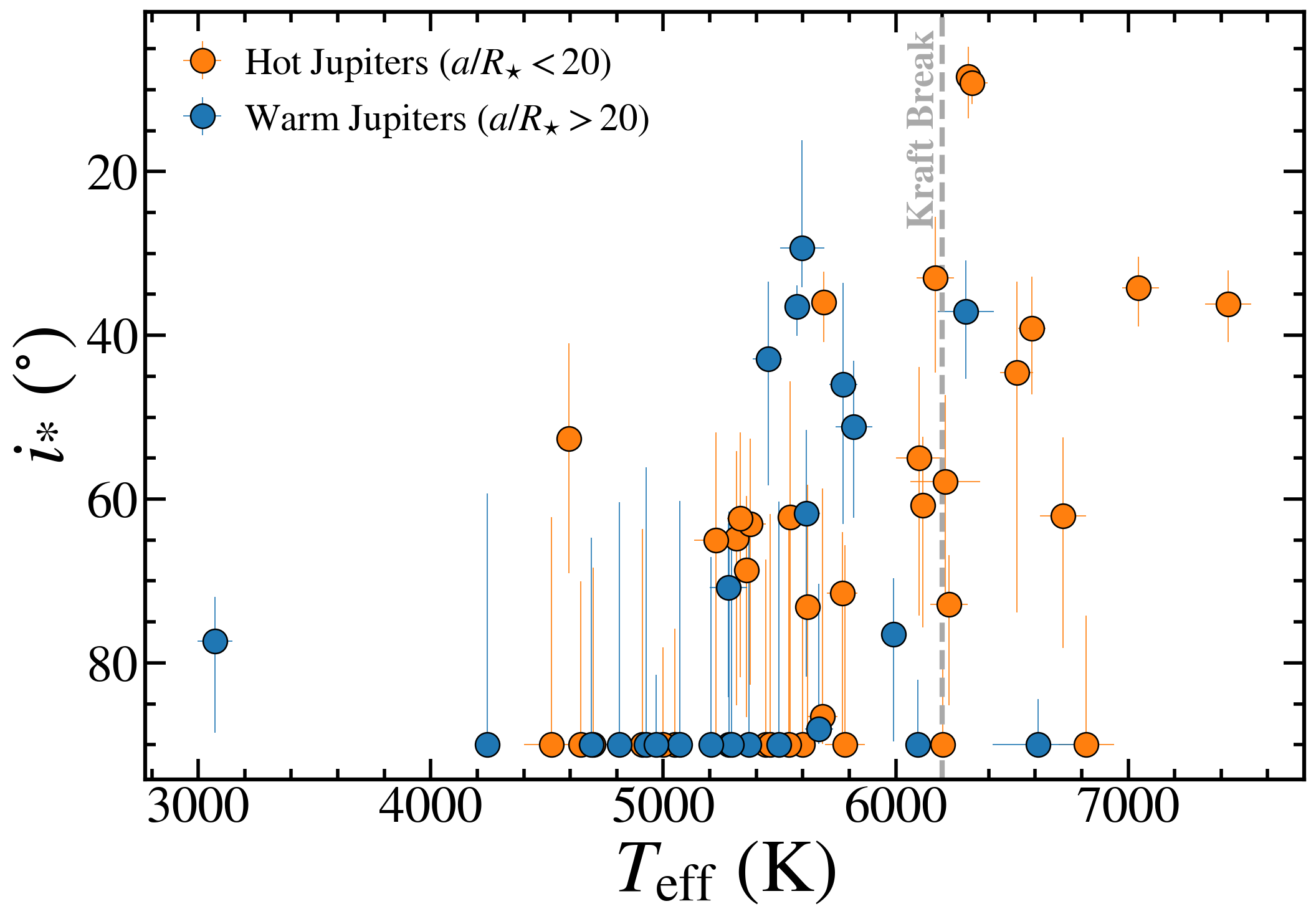}
  \hfill
    \centering
    \includegraphics[width=\linewidth]{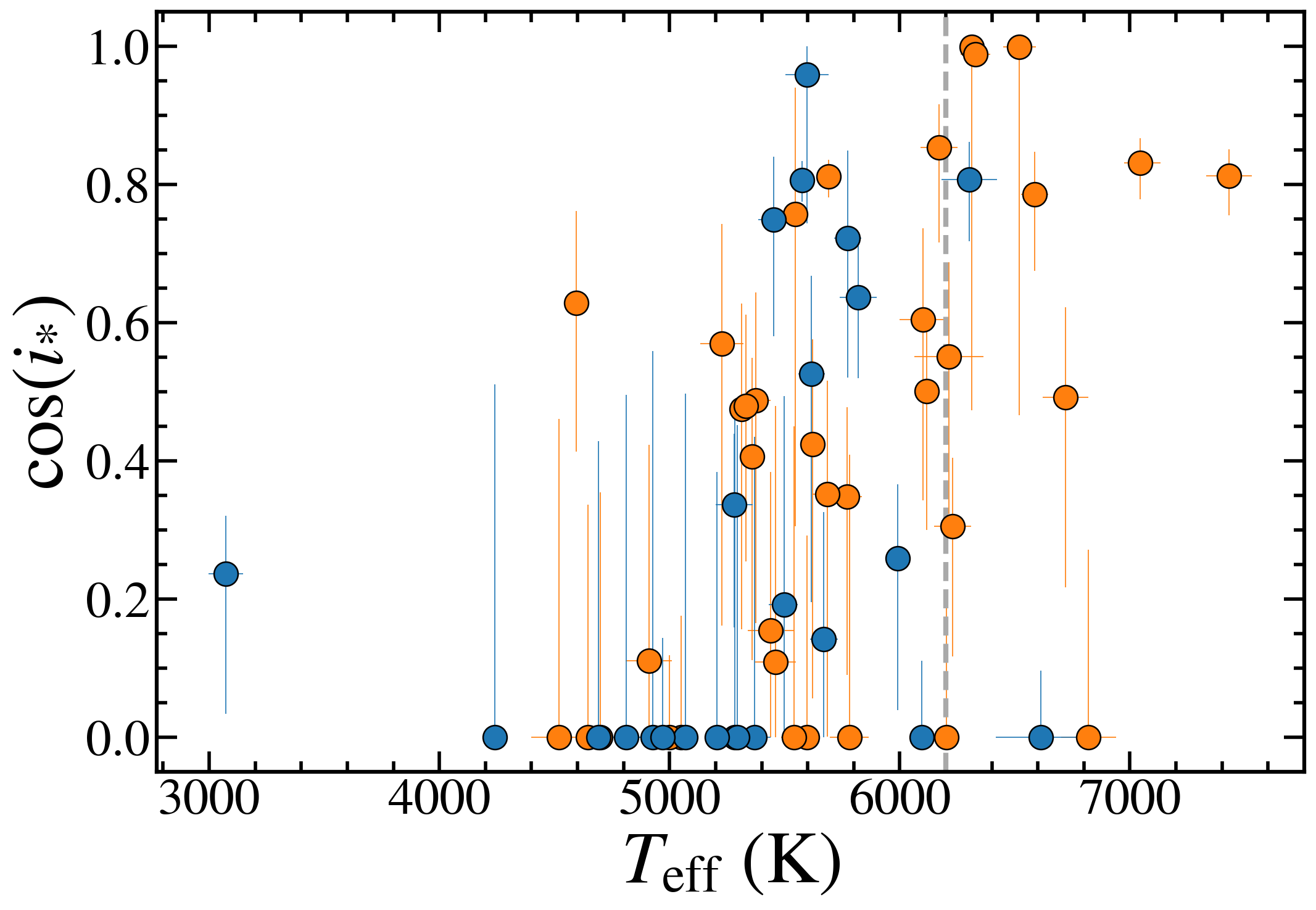}
    \caption{Top: $i_{*}$ as a function of stellar effective temperature for all stars in our sample. Bottom: cos($i_{*}$) as a function of stellar effective temperature for all stars in our sample. The gray dashed line marks the transition region above and below the Kraft break at $T_\mathrm{eff}$ $\sim$6200 K.}
  \label{{fig:kraft_break}}
  \end{minipage}
\end{figure}

\begin{figure}
\begin{center}
{\includegraphics[width=\linewidth]{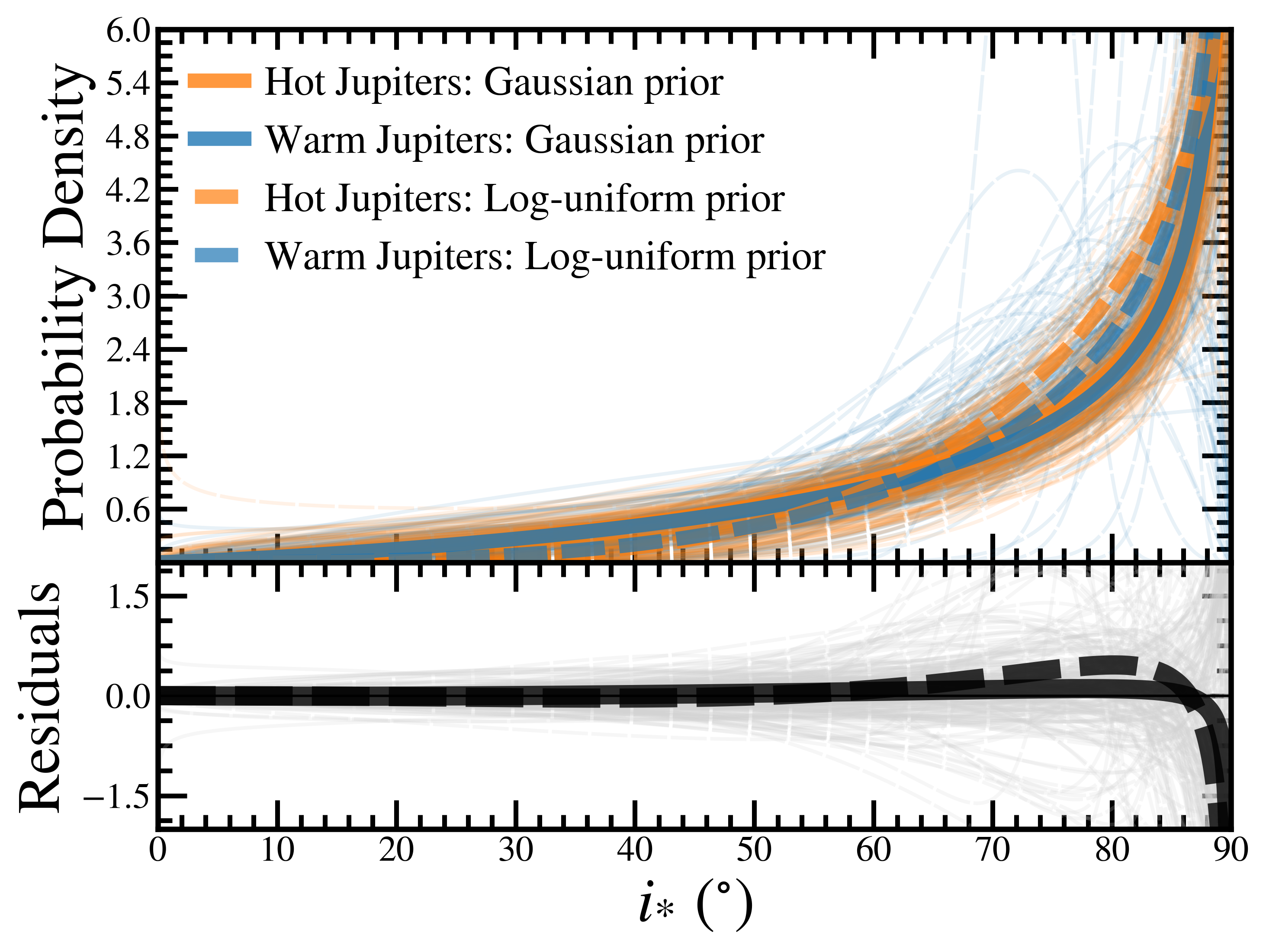}}
\caption{Results from our heirarchical Bayesian modeling of the underlying stellar inclination distributions for hot Jupiters (orange) and warm Jupiters (blue). Thin lines in the background show randomly sampled distributions from the $\alpha$ and $\beta$ hypterparameter posteriors, which represent the shape parameters of the Beta distribution. The thick orange and blue curves display the median inferred underlying $i_{*}$ distribution among all hot and warm Jupiter host stars below the Kraft break. Solid lines represent a Truncated Gaussian hyperpiror while the dashed lines represent a log-uniform hyperprior. The lower panel displays the residuals of hot Jupiter host stars subtracted from stars harboring warm Jupiters. The two underlying $i_{*}$ distributions are mutually consistent when modeled with the Beta distribution, irrespective of the different hyperpriors we tested.}
\label{fig:posterior_analysis}
\end{center}
\end{figure}

\begin{deluxetable}{lcccc}
\renewcommand\arraystretch{0.9}
\tabletypesize{\small}
\setlength{ \tabcolsep } {.1cm}
\tablewidth{0pt}
\tablecolumns{5}
\tablecaption{Beta distribution Model Posteriors from MCMC Fitting \label{tab:Model_Parameters}}
\tablehead{
\colhead{Sample} & \colhead{Hyperprior} & \colhead{$\alpha$} & \colhead{$\beta$} 
}
\startdata
    Hot Jupiter & Gaussian &  2.40$^{+0.67}_{-0.59}$  & 0.59$^{+0.24}_{-0.16}$ \\
    Warm Jupiter & Gaussian &  2.04$^{+0.63}_{-0.54}$ &  0.56$^{+0.23}_{-0.16}$\\
    Hot Jupiter & Log-uniform &  5.14$^{+4.04}_{-2.07}$ & 0.99$^{+1.09}_{-0.42}$\\
    Warm Jupiter & Log-uniform & 2.74$^{+1.43}_{-0.98}$ & 0.63$^{+0.39}_{-0.21}$ \\
\enddata

\end{deluxetable}


\section{Discussion}{\label{sec:Discussion}}

\begin{figure*}
\begin{center}
{\includegraphics[width=\linewidth]{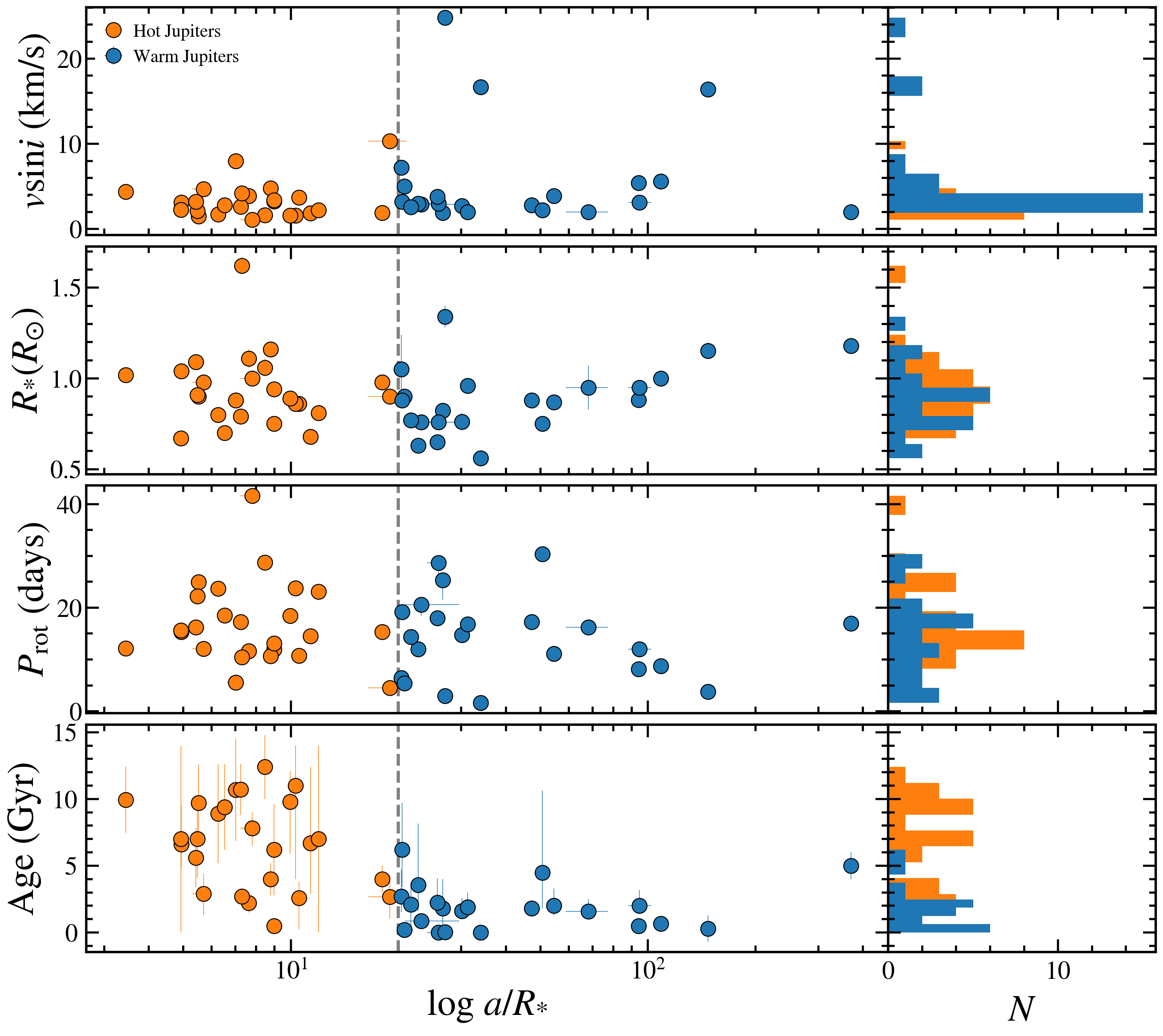}}
\caption{Stellar properties for the hot Jupiter (orange) and warm Jupiter (blue) samples as a function of scaled orbital distance, $a/R_{*}$. Marginalized histograms of each parameter are shown on the right. From top to bottom panels, projected rotational velocities, stellar radii, rotation period, and age are plotted as a function of $a/R_{*}$.  In general the hot and warm Jupiter samples show similar characteristics, but hot Jupiters have a broader spread of ages compared to warm Jupiter hosts, which are typically $<$5 Gyr. We note that K2-281 does not have a reported age and was therefore not included in the bottom panel.}
\label{fig:paramater_comparison}
\end{center}
\end{figure*}

 Only a few previous studies have attempted to compare the spin-orbit patterns of warm Jupiters to those of hot Jupiters using fully or partially constrained obliquity angles. \citet{Albrecht2012} examined the obliquities of short-period giant planet host stars and used a threshold of $a/R_{*} >$ 10 to distinguish hot and warm Jupiters. In their sample, three of the four systems below $\sim$6200 K, with large scaled orbital distances beyond $a/R_{*} >$ 10, are significantly misaligned. From this they speculated that warm Jupiters may have higher rates of misalignment compared to hot Jupiters. In this scenario, the difference between preferentially aligned hot Jupiter orientations and the broader warm Jupiter stellar obliquity distribution was attributed to tidal interactions at close separations. 
 
 \citet{2022AJ....164..104R} also compared cool stars hosting hot and warm Jupiters in a consistent fashion using a cutoff of $a/R_{*} >$ 11 to separate the two populations of planets. They found that all 12 of their warm Jupiter host stars are aligned and concluded that warm Jupiter hosts are \emph{more} aligned than their hot Jupiter counterparts at the 3.3$\sigma$ level. This suggests marginally significant evidence for a difference. This finding is counter to the trends hinted at in \citet{Albrecht2012}.

Our analysis in this study comprises 22 warm Jupiters with (minimum) obliquity constraints beyond $a/R_{*} >$ 20, making it over 5 times larger than previous samples from \citet{Albrecht2012} and \citet{2022AJ....164..104R} at the same scaled distance. We find that warm Jupiter obliquities fall between these previous results: hot and warm Jupiter host stars show a modest fraction (14--24$\%$)  of misalignments below the Kraft Break, and the underlying minimum obliquity distributions are identical, at least to within the precision available given existing sample sizes. As seen in Figure \ref{fig:posterior_analysis}, there is no discernible distinction in the underlying parent distributions of $i_{*}$ values. Furthermore, the log-uniform and truncated Gaussian hyperpriors produce underlying $i_{*}$ distributions that are similar between both populations of host stars, indicating that the posterior shapes are being driven by the data and not the hyperpriors. This is also reflected quantitatively in the consistent constraints on the Beta Distribution model parameters $\alpha$ and $\beta$ (Table \ref{tab:Model_Parameters}). This suggests that warm Jupiters are occasionally misaligned at similar rates as hot Jupiters. Warm Jupiters do not appear to have more excited obliquities (\citealt{Albrecht2012}) or more aligned obliquities (\citealt{2022AJ....164..104R}) compared to close-in giant planets.

Tidal torques have traditionally been invoked to damp stellar obliquities and realign planets at close separations (\citealt{Winn2010}; \citealt{Albrecht2012}; \citealt{Spalding2022}), but the similar minimum obliquity distributions within and beyond 0.1 AU call into question whether tides are in fact impacting the population-level obliquities of hot Jupiters. Tidal models have successfully shown that obliquity damping is possible for planets with orbital periods $\lesssim$ 3 days, however, it is less efficient at larger scaled distances (\citealt{Anderson2021}).  The small fraction of misalignments observed among host stars below the Kraft break could instead be interpreted as cool stars being set by a primordial misaligned disk distribution---mostly aligned, but with occasional misalignment. Obliquities may also occur if broken or misaligned disks torque the spin axis of the host star (\citealt{Epstein-Martin2022}).  This would imply that the hot and warm Jupiters around these cool stars either formed from coplanar planet-planet scattering or disk migration. \citet{BarkerOgilvie2009} found the timescale for spin-orbit alignment is comparable to the orbit decay time for hot Jupiters. Therefore planets that are observed to be aligned most likely formed co-planar and did not experience consequential tidal interactions. Tidal realignment would therefore not need to be introduced as the host stars are already primordially aligned or misaligned.

If tides are not dominating the obliquity distribution of hot Jupiters, the different obliquity distributions for low- and high-mass stars could instead reflect differences in the distribution of primordial misaligned disks, or some other mechanism that is preferentially exciting hot Jupiter inclinations around hot stars. For instance, high-mass stars harbor more giant planets (\citealt{Johnson2007}; \citealt{Bowler2010}; \citealt{Johnson2010}) which could increase the chances of scattering events and excite mutual inclinations. The observation that this occurs near the Kraft break could be coincidental. Indeed, \citet{HamerSchlaufman2022} recently found that stellar mass, as opposed to stellar effective temperature, is a better predictor of stellar obliquity.

Alternatively, hot Jupiter host stars may have formed with broad obliquities and over time became tidally realigned as a result of interactions bewteen the planet and the convective envelopes of cool stars, while hot stars retained a range of misalignments. In this scenario, which follows the traditional interpretation of hot Jupiter obliquity patterns, warm Jupiters around cool stars are predominantly formed aligned and migrate through mechanisms that do not excite inclinations, like coplanar planet-planet scattering or disk migration. However, for this to hold true, we would have to be observing the realignment process of hot Jupiters at a special time where it happens to be consistent with the warm Jupiter host star obliquity distribution. In the past, the hot Jupiter distribution would have been much broader and in the future (several Gyr from now) it would be narrower.

In summary, we conclude that the consistency of alignment between both hot and warm Jupiter host stars indicates that either tidal realignment is not shaping the hot Jupiter obliquity distribution, or we are observing the hot Jupiter realignment process at a time that happens to match the obliquity distribution of warm Jupiters.

\begin{figure}
\begin{center}
{\includegraphics[width=\linewidth]{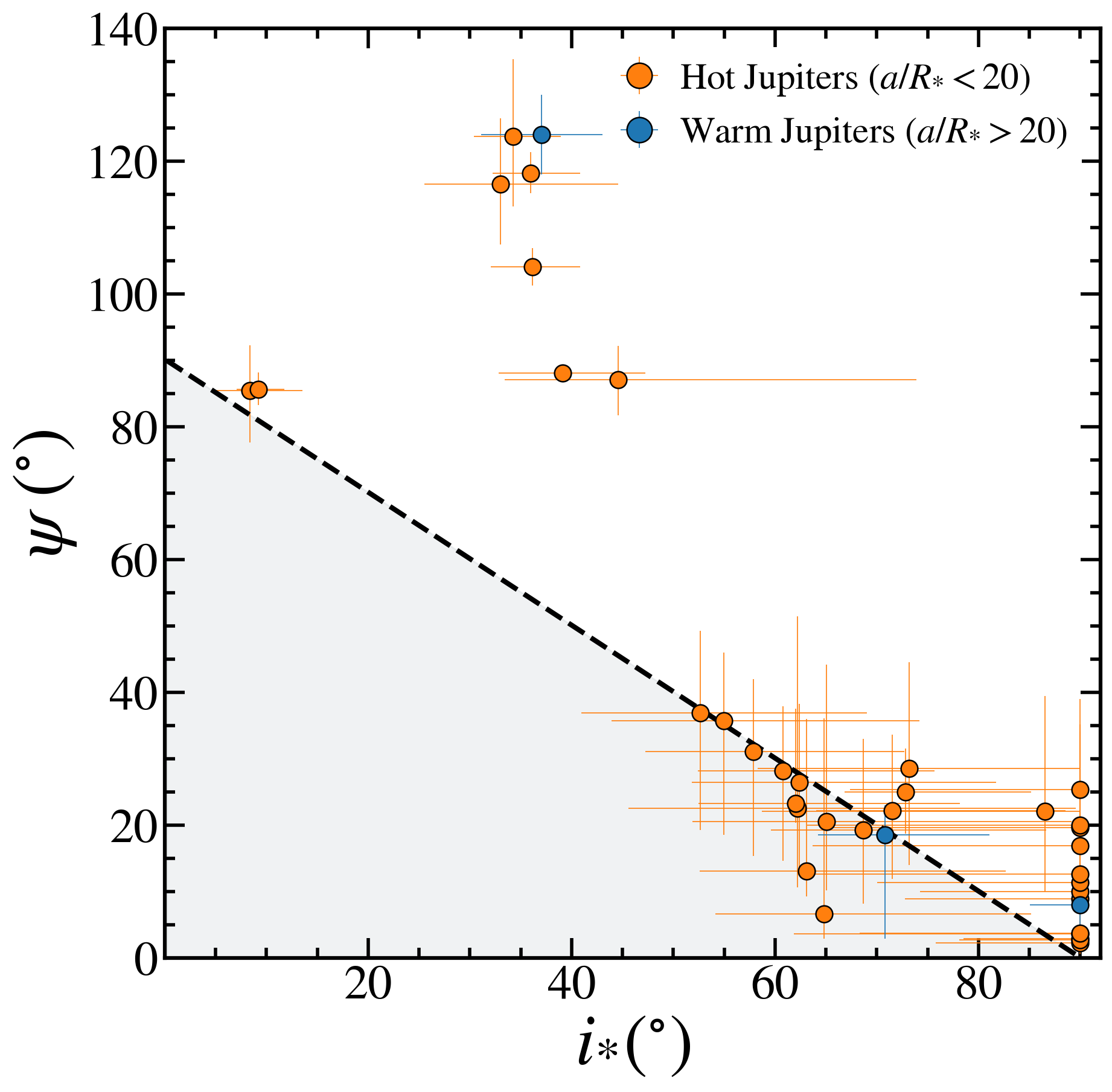}}
\caption{$\psi$ as a function of $i_{*}$ for all systems with measurements in our sample (\citealt{2022PASP..134h2001A}; \citealt{Johnson2022}). We find stars that are most likely misaligned in $i_{*}$ space are also most likely misaligned in $\psi$ space as well. This generates a lower limit on the inferred misalignment. The dashed line corresponds to minimum misaligned obliquities and represents the case where $\lambda$ = 0, so cos($\psi$) = sin($i_{*}$). Any departure from 0 will increase $\psi$ for a given $i_{*}$. The dashed curve corresponds to minimum misaligned obliquities and represents the case where $\lambda$ = 0, so cos($\psi$) = sin($i_{*}$).}
\label{fig:minimum_misalignment}
\end{center}
\end{figure}

\begin{figure}
\begin{center}
{\includegraphics[width=\linewidth]{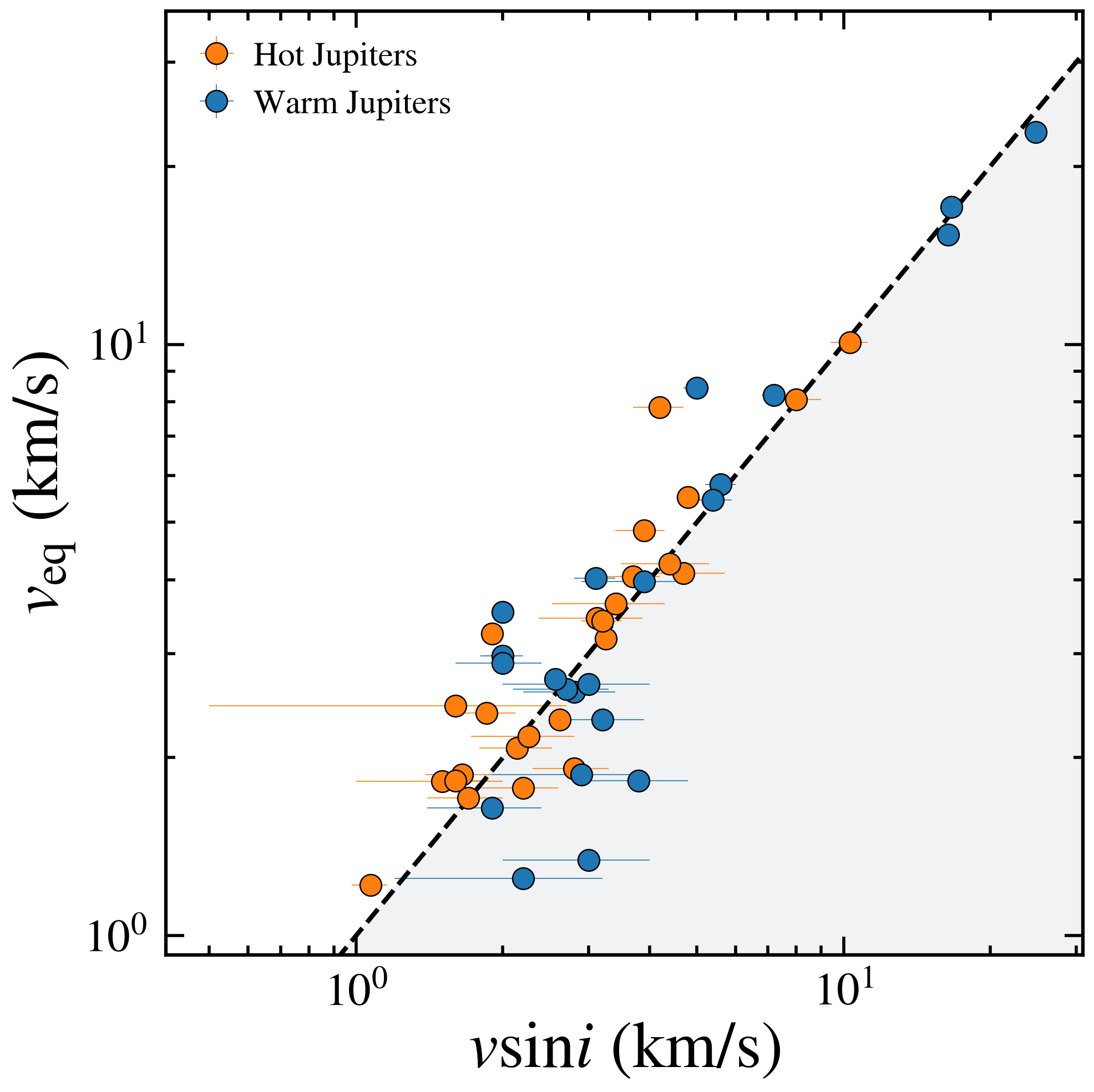}}
\caption{$v \sin i$ plotted as a function of $v_\mathrm{eq}$ for all systems below the Kraft break in our sample. The dashed line represents where our measured $v_\mathrm{eq}$ values are in agreement with adopted $v \sin i$ values. The gray shaded area corresponds to an unphysical region in which the projected rotational velocity is larger than the equatorial velocity. All stars have $v \sin i$ values that are within 2$\sigma$ of their equatorial velocities. Values to the upper left of the 1:1 line represent misaligned stellar obliquities.}
\label{fig:vsin_veq}
\end{center}
\end{figure}

It is also interesting to consider the broader obliquity distribution of cold Jupiters at wider separations. Little is known about spin-orbit angles for giant planets beyond  $\sim$2 AU. However, \citet{Bowler2023} report that stars hosting directly imaged planets within 20 AU mostly show angular momentum alignment, in contrast to more massive brown dwarf companions. The trend of low obliquities for warm Jupiters may therefore extend to wide separations, although the sample of imaged planets with obliquity constraints remains quite limited (\citealt{Kraus2020}).

Future studies and additional observations are needed to distinguish which migration channels are dominating the hot and warm Jupiter populations. To further assess primordial misalignments and planet-host star interactions, observations of hot Jupiters around young stars, warm Jupiters around hot stars, and the primordial distribution of protoplanetary disk orientations would be helpful. Although each of these tests would be informative, they possess their own significant observational challenges. For instance, there are few hot Jupiters known around young stars, and it is difficult to measure obliqities of hot stars harbouring warm Jupiters. We did not include RM measurements in this analysis, but measurements of the projected angle between the orbital and stellar spin axes $\lambda$, combined with the stellar inclination $i_{*}$, will be valuable to fully characterize full obliquity measurements to these systems (\citealt{Albrecht2012}; \citealt{2021ApJ...916L...1A}; \citealt{Rice2022a}; \citealt{2022AJ....164..104R}).

\subsection{Potential Biases}\label{sec:potential_biases}

\begin{deluxetable*}{lcccccc}
\renewcommand\arraystretch{0.9}
\tabletypesize{\small}
\setlength{ \tabcolsep } {.1cm}
\tablewidth{0pt}
\tablecolumns{5}
\tablecaption{Misalignment Probabilities \label{tab:misalignment_percentiles}}
\tablehead{
    & \colhead{Misalignment Threshold} & \multicolumn{4}{c}{Probability Threshold} \\
  \colhead{Sample} & \colhead{ $\ge$ $\Delta i$} & \multicolumn{2}{c}{$>$ 80$\%$} & \multicolumn{2}{c}{$\ge$ 90$\%$} }
\startdata
            Hot Jupiter & 5$^\circ$ & $23/25$ & 89$^{+3}_{-6}\%$  & $12/25$ & 50$^{+9}_{-9}\%$ \\
            Warm Jupiter & 5$^\circ$ & $20/22$ & 89$^{+5}_{-7}\%$ & $6/22$ & 29$^{+10}_{-8}\%$\\
            Hot Jupiter & 10$^\circ$ & $9/25$ & 37$^{+10}_{-8}\%$ & $3/25$  & 14$^{+7}_{-5}\%$\\
            Warm Jupiter & 10$^\circ$ & $6/22$ & 29$^{+10}_{-8}\%$ & $5/22$ & 24$^{+9}_{-7}\%$ \\
            Hot Jupiter & 20$^\circ$ & $2/25$& 10$^{+6}_{-3}\%$ & $2/25$ & 10$^{+6}_{-3}\%$\\
            Warm Jupiter & 20$^\circ$ & $5/22$ & 24$^{+9}_{-7}\%$ & $2/22$ & 12$^{+7}_{-4}\%$\\
\enddata
\end{deluxetable*}
		
Here we outline potential biases in this analysis.  These could in principle impact our results, either in an absolute sense (such as by biasing measurements) or in a relative sense (for instance, when comparing hot and warm Jupiter distributions).  We argue that while there are several ways to individually bias $i_*$ values or $i_*$ distributions, it is unlikely that these impact the relative comparison of hot and warm Jupiter obliquities---a key result from this study.

\subsubsection{i* Analysis Bias}\label{sec:i_star_analysis_bias}

Our results rely on the homogeneous and self-consistent analysis of $i_{*}$ measurements. These measurements provide meaningful constraints on misalignments, but without sky-projected obliquities (through RM measurements, for instance), the true obliquity cannot be fully determined. There are several factors that could bias stellar inclinations including overestimated $v \sin i$ values for slow rotators, miscalculation of rotation periods due to spots at non-equitorial latitudes, and over (or under) estimates of $R_{*}$ from evolutionary models or SED fitting. However, when comparing the distributions of these parameters for hot and warm Jupiters, there are no indications of strong differences that might impact one sample over the other.  

\subsubsection{Age Bias}\label{sec:age_bias}
The formation of hot Jupiters from disk migration or high-eccentricity migration can occur over a broad range of timescales from a disk lifetime to a Hubble time. However, once hot Jupiters have migrated, tidal realignment is generally expected to operate on long timescales of several Gyr for planets with orbital periods greater than a few days (\citealt{RasioTides1996}; \citealt{Spalding2022}).  Age is therefore an important parameter to consider between our hot and warm Jupiter samples.  If tidal torques shape the hot Jupiter obliquity distribution, then a young hot Jupiter sample might show a broader $i_*$ distribution while an older population would be preferentially aligned. This could impact the interpretation of our comparison of the reconstructed stellar inclination distributions.

\begin{figure}
\begin{center}
{\includegraphics[width=\linewidth]{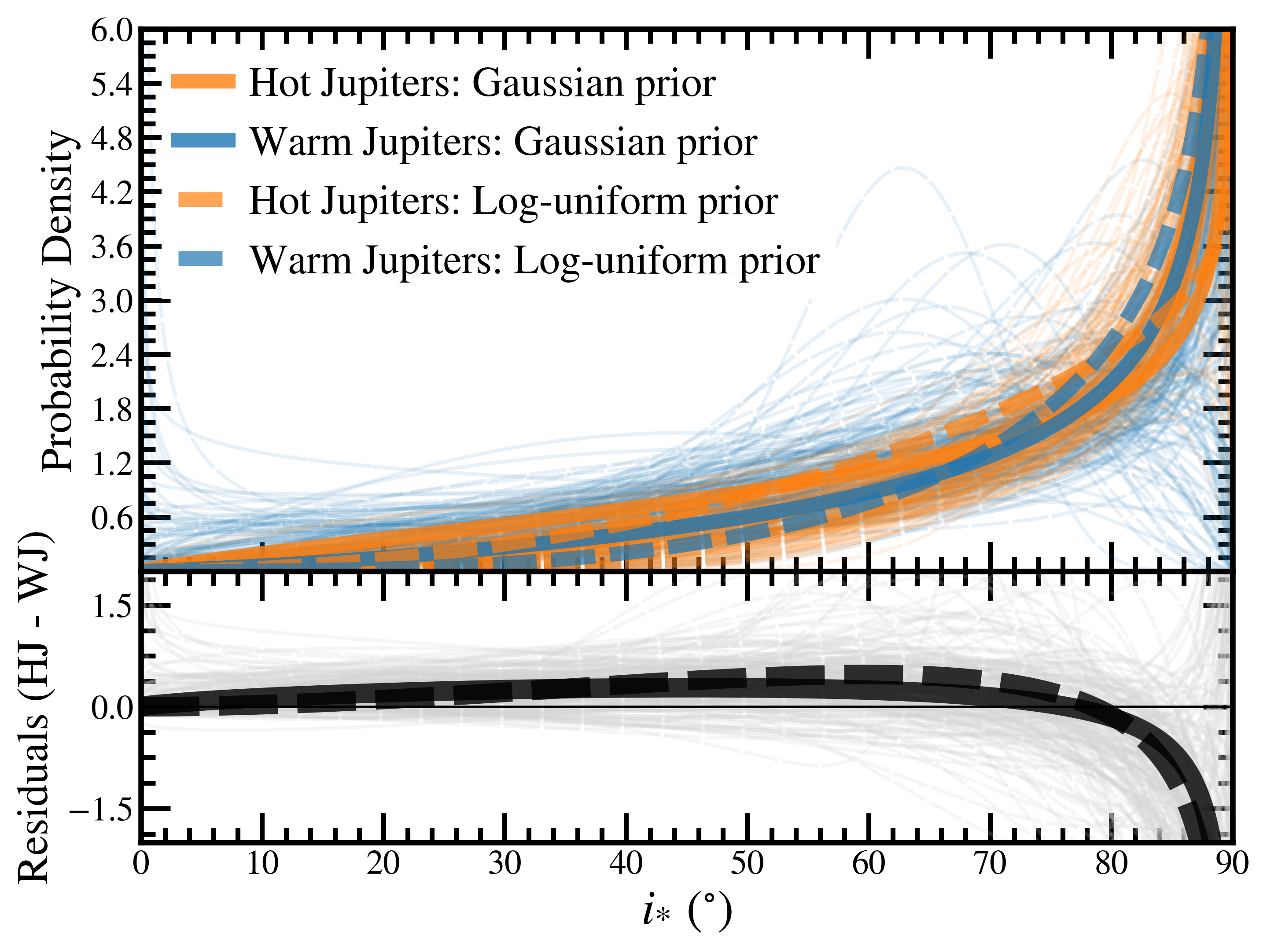}}
\caption{Results from our hierarchical Bayesian modeling test focusing on age as a potential bias. Underlying stellar inclination distributions for hot and warm Jupiter host stars younger than 6 Gyr are displayed. See Figure \ref{fig:posterior_analysis} for details.}
\label{fig:age_analysis}
\end{center}
\end{figure}

In Figure \ref{fig:paramater_comparison}, we find that the warm Jupiter population is on average younger than the hot Jupiter population. The hot Jupiter sample spans all ages while the warm Jupiters only have ages up to 6 Gyr. One explanation for the younger warm Jupiter sample is that we only include systems with readily retrievable light curve rotation periods. This biases our warm Jupiter population to younger systems because rotation periods are shorter, starspot covering fractions are larger, and the light curve amplitudes are higher. To assess the impact of the broader hot Jupiter ages, we ran additional Hierarchical Bayesian statistical tests to more fairly compare the hot and warm Jupiter populations by selecting systems with effective temperatures less than 6200 K and younger than 6 Gyr (the full range of the warm Jupiter sample). For the hot Jupiter sample $<$ 6 Gyr, we find $\alpha$ and $\beta$ values of [1.84$^{+0.7}_{-0.60}$, 0.76$^{+0.46}_{-0.31}$] and  [4.06$^{+6.37}_{-2.25}$, 1.47$^{+3.38}_{-0.91}$] for Gaussian and log-uniform hyperpriors, respectively. The warm Jupiter sample $<$ 6 Gyr, yields $\alpha$ and $\beta$ values of [2.04$^{+0.63}_{-0.54}$, 0.57$^{+0.23}_{-0.16}$] and [2.74$^{+1.43}_{-0.97}$, 0.63$^{+0.38}_{-0.21}$] for Gaussian and log-uniform priors, respectively. There is no significant difference in the underlying frequency of host star alignment or reconstructed $i_{*}$ distributions, as seen in Figure \ref{fig:age_analysis}. We conclude that the broader age distributions of hot Jupiters does not appear to impact the results or interpretation from this work. 

\subsubsection{Orbital Distance Bias}\label{sec:orbital_distance_bias}
\begin{figure}
\begin{center}
{\includegraphics[width=\linewidth]{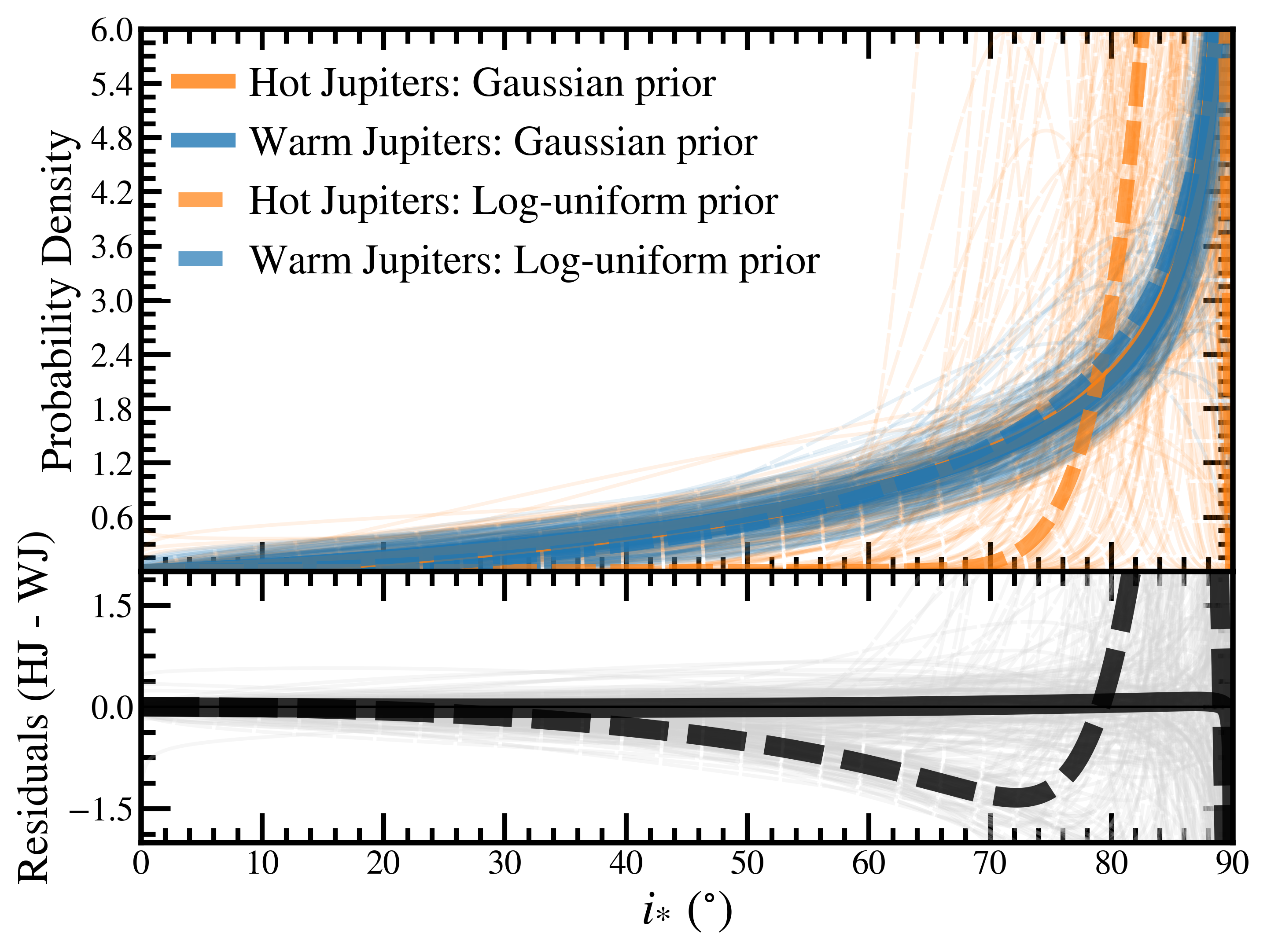}}
\caption{Results from our hierarchical Bayesian modeling test focusing on scaled orbital distance as a potential bias. Underlying stellar inclination distributions with an $a/R_{*}$ cut at 10 for hot and warm Jupiter host stars are displayed. See Figure \ref{fig:posterior_analysis} for details.}
\label{fig:a_R_analysis}
\end{center}
\end{figure}

It is also possible that our choice for scaled orbital distance to separate the hot and warm Jupiter samples could impact the results, especially given the modest sizes of both samples. To test this, we account for orbital distance as a potential bias by separating the hot and warm Jupiter populations with an $a/R_{*}$ cut at 10. This value is closer to the distinction of the two Jovian populations in \citet{Albrecht2012} and \citet{2022AJ....164..104R}. Additional HBM statistical tests are run with an $a/R_{*}$ cut of 10 and an effective temperature cut at 6200 K to isolate cool host stars. For the hot Jupiter sample ($a/R_{*}$ $<$ 10), we find $\alpha$ and $\beta$ values of [2.36$^{+0.71}_{-0.63}$, 0.54$^{+0.24}_{-0.16}$] and [31.74$^{+39.87}_{-21.81}$, 1.55$^{+3.56}_{-0.87}$] for Gaussian and log-uniform priors, respectively. We note that with this particular test, the hot Jupiter sub-sample is more prior dependent than other tests we ran. The warm Jupiter sample ($a/R_{*}$ $>$ 10), produced $\alpha$ and $\beta$ values of [2.05$^{+0.63}_{-0.54}$, 0.56$^{+0.23}_{-0.16}$] and [3.14$^{+1.41}_{-1.01}$, 0.61$^{+0.29}_{-0.18}$] for Gaussian and log-uniform priors, respectively. No distinct difference is evident when compared to our nominal threshold of $a/R_{*}$ = 20 as seen in Figure \ref{fig:a_R_analysis}.  We conclude that our specific choice of $a/R_{*}$ to define the hot and warm Jupiter samples does not appear to impact the results.

\subsubsection{Small Sample Bias}\label{sec:small_sample_bias}
\citet{Bowler2020} and \cite{Nagpal2022} performed tests to assess how reliably an input underlying distribution could be reproduced using HBM with simulated measurements as a function of sample size and measurement uncertainty.  Although their experiments were carried out for eccentricities, the results can equally apply to stellar inclinations.  Several hyperpriors on the Beta distribution shape parameters were tested; \cite{Nagpal2022} found that a Truncated Gaussian hyperprior reliably recovered the characteristic shape of the input distribution for sample sizes as small as 5 and eccentricity uncertainties as large as 0.2, which corresponds to stellar inclination uncertainties of 18$\degr$.  Samples of 20---similar to the sizes used in this study---were even more accurate and substantially improved the precision of the posterior distribution.  This indicates that although the samples remain modest for  the hot and warm Jupiter populations, the consistency of the recovered underlying distributions is expected to be robust.

\subsubsection{Viewing Angle Bias}\label{sec:viewing_angle_bias}
In order to infer true spin-orbit angles, the sky-projected obliquity, stellar inclination, and inclination of the planet's orbital plane must be known. The most readily way to fix one of these parameters is through an edge-on configuration with transiting planets. However, if $i_{*}$ is known, the orientation of the spin axis relative to the orientation of the orbital plane implies that $i_{*}$ is a minimum misalignment angle, and bounds $\psi$ to be between $|$ $\Delta$$i$ $|$ $\approx$ $|$ $\pi$ - $i_{*}$ $|$ (for a transiting planet) and $\approx$ $\pi$ -- $|$ $\pi$ -- $i_{*}$ $|$ 
or between 0$\degr$ and 180$\degr$.  This means that true obliquities can, in principle, be very different than inferred minimum obliquities, even for large values of $i_{*}$.    

Figure \ref{fig:minimum_misalignment} demonstrates how $\psi$ values from the literature correlate with our measurements of $i_{*}$ for systems in our sample where true obliquities are available. A lower $i_*$ and higher $\psi$ measurement both indicate increased misalignment. All systems consistent with misalignment in our $i_{*}$ analysis are also misaligned in $\psi$ space. This demonstrates that our constraints on $i_{*}$ (as well as $P_\mathrm{rot}$, $v \sin i$, and $R_{*}$) are reasonable as they do not fall below the 1:1 relation when compared with $\psi$.  It also illustrates that while $\psi$ can and does depart from $i_{*}$, a preferentially aligned distribution in $\lambda$, like that of the hot Jupiters around cool stars, also imprints a preferentially aligned distribution in $i_{*}$. In addition, in Figure \ref{fig:vsin_veq} we show that our equatorial velocities are consistent to within 2 sigma of our adopted  $v \sin i$ measurements. This further reinforces the reliability of our $P_\mathrm{rot}$, $v \sin i$, and $R_{*}$ measurements.

Differentiating between co-planirity and the alignment of the star's rotational and planet's orbital axis could also play a role. For instance, the star and planet could both be coplanar but the planet could be orbiting retrograde. This could impact both our true reconstructed underlying $i_{*}$ distributions and our relative comparison, if there is a significant difference between the rate of hot and warm Jupiters on retrograde orbits. One argument against this playing a significant role comes from RM measurements where the fraction of retrograde orbits is small (\citealt{Triaud2010}; \citealt{Albrecht2012}; \citealt{2022PASP..134h2001A}). Most hot Jupiters with RM measurements have been found to orbit prograde, although the rate of retrograde warm Jupiters is not yet established.

\section{Notes on Individual Systems}\label{sec:Individual_Systems}
 With at least 90$\%$ confidence, we report 3 new misaligned transiting planets in this study based on the inferred inclination of the rotational axis of their host stars. These systems are misaligned by at least 10$^\circ$ and have MAP values of $i_{*}$ $<$ 80$^\circ$.

\subsection{Kepler-1654}\label{sec:Kepler-1654}
Kepler-1654 is a G-type star hosting a 0.8 $R_\mathrm{Jup}$ planet on a 1047-day orbit (2.0 AU) (\citealt{2018AJ....155..158B}).  We report a line-of-sight stellar spin inclination of 29$^{+5}_{-13}\degr$. Here we have adopted the upper limit of 2 km s$^{-1}$ from \citet{Petigura2015} to infer the posterior distribution of $i_{*}$, which is shown in Figure \ref{fig:obliquities_pg3}. This implies that the star’s equatorial plane is significantly misaligned with the orbital plane of the planet by at least 61$^{+13}_{-5}\degr$. To date, Kepler 1654 b is the longest-period giant planet found in a misaligned system. The origin of the misalignment may be a sign of an undetected planetary companion.

\subsection{Kepler-539}\label{sec:Kepler-539}
Kepler-539, a solar-type G star hosting a giant planet with a minimum mass of 0.97 $M_\mathrm{Jup}$ and a period of 125 days (0.5 AU), was announced by \citet{Mancini2016}. We derive a stellar inclination of $i_{*}$ = 51$^{+11}_{-8}\degr$. The full posterior distribution is shown in Figure \ref{fig:obliquities_pg3}. After Kepler-1654 b, Kepler-539 b is the second-longest orbiting planet in a misaligned system currently known, with a misalignment between the orbital plane and stellar inclination of 39$^{+8}_{-11}\degr$.

\subsection{Kepler-30}\label{sec:Kepler-30}
 Kepler-30 is a Sun-like star that hosts three transiting planets (\citealt{Fabrycky2012}). Kepler-30 c is a giant planet in this multi-planet system with a minimum mass of 2 $M_\mathrm{Jup}$ and period of 60 days (0.3 AU). Using two different methods to measure $\lambda$, \citet{Sanchis-Ojeda2012} report values of 4$^{+10}_{-10}\degr$ and -1$^{+10}_{-10}\degr$, suggesting alignment of the stellar spin axis with the orbital plane. We derive an $i_{*}$ = 43$^{+15}_{-9}\degr$, as can be seen in Figure \ref{fig:obliquities_pg2}. We combine the two independent measurements of $\lambda$ from \citet{Sanchis-Ojeda2012} as a weighted mean ($\lambda$ = 1.5 $\pm$ 7.1$\degr$) together with $i_*$ and report a 3D obliquity $\psi$ = 90$^{+34}_{-34}\degr$. This differs from the conclusions drawn by \citet{Sanchis-Ojeda2012} because although the projected stellar spin axis may be aligned with the orbital plane, the stellar inclination is misaligned, resulting in a high obliquity. This severe offset suggests the misalignment was present during the system's early stages when planets were forming in the disk, or some other mechanism has subsequently tilted the star's orientation after formation. The Kepler-30 system joins other misaligned multi-planet systems with $\psi$ $\gtrsim$ 40$\degr$ including K2-290 (\citealt{Hjorth2019}; \citealt{Hjorth2021PNAS}; \citealt{Best2022}), Kepler-56 (\citealt{Huber2013}; \citealt{Otor2016}), and Kepler-129 (\citealt{Zhang2021}).

\section{Conclusion}\label{sec:Conclusion}
In this work, we presented line-of-sight inclinations for 48 cool stars harboring giant planets. We find Kepler-1654 b and Kepler-539 b to be two of the longest-period giant planets known in misaligned systems. In addition, Kepler-30 is a newly identified misaligned multi-planet system. By comparing the reconstructed underlying $i_{*}$ distributions using heirarchical Bayesian modeling, we do not find a distinct difference between the inferred minimum misalignments of hot and warm Jupiter host stars. Below the Kraft break, we find with $90\%$ confidence that $24^{+9}_{-7}\%$ of warm Jupiters and $14^{+7}_{-5}\%$ of hot Jupiters are misaligned by at least $i_{*}$ = 10$^\circ$. 

There are two broad interpretations when considering this result together with the excited obliquity distribution of hot Jupiters around hot stars. 
\begin{itemize}[topsep=0pt, partopsep=0pt, itemsep=0pt, parsep=0pt]
\item In the first scenario, giant planets form and undergo inward coplanar migration in aligned disks. Tidal realignment of hot Jupiters is not damping the obliquity distribution around cool stars, and instead the obliquities of more massive stars are preferentially excited, perhaps because of increased scattering in the presence of more multiple giant planets or a broader initial disk distribution. The transition near the Kraft Break is coincidental and the differences in obliquity distributions between cool and hot stars is best described by stellar mass. 
\item Alternatively, tidal realignment is operating, but the evolving hot Jupiter obliquity distribution (from broad to narrow) happens to match the warm Jupiter distribution right now at the typical ages of field stars. 

\end{itemize}

Further observations of transiting planets around evolved stars, hot Jupiters around young stars, warm Jupiters around hot stars, and the distribution of protoplanetary disk orientations will be necessary to disentangle the primordial and post-formation misalignment hypotheses. Obliquity measurements will provide valuable clues into the relatively unknown dynamical histories of these two planet populations.

\section{acknowledgments}
We thank Rebekah Dawson, Eugene Chiang, and J.J. Zanazzi for insightful conversations. B.P.B. acknowledges support from the National Science Foundation grant AST-1909209, NASA Exoplanet Research Program grant 20-XRP20$\_$2-0119, and the Alfred P. Sloan Foundation.

This research has made use of the NASA Exoplanet Archive, which is operated by the California Institute of Technology, under contract with the National Aeronautics and Space Administration under the Exoplanet Exploration Program and \citet{VizieR2000}, an online database with sources collected by the Centre de Données de Strasbourg (CDS).

\software{\texttt{numpy} \citep{Harris2020}, \texttt{matplotlib}  \citep{Hunter2007}, \texttt{ePop!} \citep{Nagpal2022}, \texttt{corner} \citep{Foreman-Mackey2016}, \texttt{emcee} \citep{Foreman-Mackey2013}, and \texttt{pandas} \citep{reback2020pandas}}

\begin{longrotatetable}
\begin{deluxetable*} {lcccccccccccccc}
\centerwidetable
\renewcommand\arraystretch{0.95}
\setlength{\tabcolsep}{.2cm}
\tabletypesize{\small}
\tablecaption{Host star properties.\label{tab:host_stars}}
\tablehead{
     \colhead{Name} & \colhead{Light Curve} & \colhead{$P_\mathrm{rot}$} & \colhead{$\sigma_{P\mathrm{,tot}}$} & \colhead{$v$sin$i_*$} & \colhead{$\sigma_{v \mathrm{sin} i}$} & \colhead{$i_*$} & \colhead{$\sigma_{i_{*}}$} & \colhead{$a/R_*$} & \colhead{$\sigma_{a/R_{*}\mathrm{,tot}}$} & \colhead{$R_*$} & \colhead{$\sigma_{R_{*}\mathrm{,tot}}$} & \colhead{$T_\mathrm{eff}$} & \colhead{Age} & \colhead{Ref.} \\
       & \colhead{Provenance} & \colhead{(d)} & \colhead{(d)} & \colhead{(km s$^{-1}$)} & \colhead{(km s$^{-1}$)} & \colhead{($^\circ$)} & \colhead{($^\circ$)} &  & \colhead{$(R_\odot)$} & \colhead{$(R_\odot)$} & & \colhead{(K)}& \colhead{(Gyr)}
        }
\startdata
\multicolumn{15}{c}{Warm Jupiters ($a$/$R_*$ $>$ 20)} \\
\hline
Kepler-9 & \textit{Kepler} Q1-17 & 16.82 & 0.06 &2.0 & 0.4 & 46.01 & $^{+17.02}_{-12.38}$ & 31.3 & 1.1 & 0.96 & 0.02 & 5774$^{+60}_{-60}$ & 1.91$^{+1.09}_{-0.57}$ & 1 \\
Kepler-27 & \textit{Kepler} Q1-17 & 14.77 & 0.06 & 2.7 & 0.6 & 90 & $^{+0.14}_{-27.28}$ & 30.12 & 0.6 & 0.761 & $^{+0.049}_{-0.046}$ & 5294$^{+68}_{-56}$	& 1.62$^{+0.98}_{-0.47}$ & 2,3 \\
Kepler-28 & \textit{Kepler} Q1-17 & 17.97 & 0.08 & 3.8 & 1.0 & 90 & $^{+0.05}_{-25.26}$ & 25.75 & 0.80 & 0.649 & $^{+0.029}_{-0.027}$ &4690$^{+63}_{-50}$	& 2.24$^{+1.82}_{-0.77}$& 2,3 \\
Kepler-30 & \textit{Kepler} Q1-17 & 16.19 & 0.06 & 2.0 & 0.2 & 42.86 & $^{+15.49}_{-9.41}$ & 68.1 & 9.2 & 0.95 & 0.12 & 5452$^{+58}_{-68}$ & 1.58$^{+0.92}_{-0.42}$ & 4,5 \\
Kepler-51 & \textit{Kepler} Q1-17 & 8.18 & 0.03 & 5.4 & 0.5 & 88.11 & $^{+1.8}_{-17.78}$ & 94.1 & 2.2 & 0.881 & 0.011 & 5670$^{+60}_{-60}$ & 0.5$^{+0.25}_{-0.25}$ & 6,7 \\
Kepler-52 & \textit{Kepler} Q1-17 & 11.99 & 0.06 & 3.0 & 1.0 & 90 & $^{+0.09}_{-30.71}$ & 23.07 & 0.37 & 0.63 & $^{+0.01}_{-0.02}$ & 4242$^{+41}_{-35}$& 3.55$^{+4.6}_{-1.95}$ & 8,9 \\
Kepler-63 & \textit{Kepler} Q0-17 & 5.4 & 0.03 & 5.0 & 0.3 & 36.51 & $^{+3.52}_{-2.65}$ & 20.79 & 0.46 & 0.9 & $^{+0.027}_{-0.022}$ & 5576$^{+50}_{-50}$	& 0.21$^{+0.05}_{-0.05}$ & 10 \\
Kepler-75  & \textit{Kepler} Q1-17 & 19.2 & 0.08 & 3.2  & 0.7 & 90 & $^{+0.09}_{-22.92}$ & 20.49 & 0.75 & 0.88 & 0.04 & 5206$^{+40}_{-45}$& 6.2$^{+3.5}_{-2.8}$ & 11,12 \\
Kepler-289 & \textit{Kepler} Q1-17 & 8.74 & 0.11 & 5.6 & 0.4 & 76.49 & $^{+13.1}_{-6.84}$ & 108.6 & 1.1 & 1.0 & 0.02 & 5990$^{+38}_{-38}$& 0.65$^{+0.44}_{-0.44}$ & 3,13 \\
Kepler-447 & \textit{Kepler}  & 6.47 & 0.03 &7.2  & 0.4 & 61.77 & $^{+19.9}_{-10.22}$ & 20.41 & $^{+0.36}_{-0.19}$ & 1.05 & 0.19 & 5615$^{+60}_{-55}$&2.69$^{+2.02}_{-1.16}$& 14 \\
Kepler-468 & \textit{Kepler} Q1-17 & 11.09 & 0.06 & 3.9 & 1.0 & 90 & $^{+0.09}_{-29.71}$ & 54.52 & 1.25 & 0.87 & $^{+0.02}_{-0.01}$ &5498$^{+60}_{-68}$	&2$^{+1.29}_{-0.73}$ & 9 \\
Kepler-470 & \textit{Kepler} Q0-17 & 24.69 & 0.14 & 20.9 & 1  & 90 &$^{+0.05}_{-5.58}$  & 24.04 & 1.082 & 1.66 & $^{+0.67}_{-0.3}$ & 6613$^{+197}_{-197}$ & 1.86$^{+0.68}_{-0.43}$  & 9 \\
Kepler-486 & \textit{Kepler} & 30.39 & 0.25 & 2.2  & 1.0 & 90 & $^{+0.05}_{-33.9}$ & 50.62 & 1.35 & 0.75 & $^{+0.02}_{-0.03}$ &4926$^{+44}_{-45}$	& 4.47$^{+6.15}_{-2.66}$& 9 \\
Kepler-539 & \textit{Kepler} Q0-17 & 11.97 & 0.03 & 3.1 & 0.3 & 51.19 & $^{+11.12}_{-8.06}$ & 94.61 & $^{+7.53}_{-6.50}$ & 0.95 & 0.02 &5820$^{+80}_{-80}$	 & 2$^{+1.2}_{-0.63}$ & 15 \\
Kepler-1654 & \textit{Kepler} Q0-17 & 16.95 & 0.08 & $<$ 2 & 0.0 & 29.35 & $^{+4.81}_{-13.18}$ & 370.3 & $^{+2.2}_{-4.7}$ & 1.18 & 0.03 & 5597$^{+95}_{-93}$  & 5$^{+1}_{-1}$ & 16 \\
K2-77 & K2 Campaign 4 & 20.56 & 2.15 & 2.9 & 1.0  & 90 & $^{+0.05}_{-29.67}$ & 23.2 & $^{+6.4}_{-2.3}$ & 0.76 & 0.03 & 5070$^{+50}_{-50}$ & 0.85$^{+0.45}_{-0.45}$& 17 \\
K2-139 & K2 Campaign 7 & 17.26 & 1.53 & 2.8 &0.6  & 90 & $^{+0.09}_{-26.11}$ & 47.25 & $^{+0.73}_{-1.98}$ & 0.88 & 0.01 & 5370$^{+68}_{-68}$	& 1.8$^{+0.3}_{-0.3}$  & 18,19 \\
K2-281 & K2 Campaign 8 & 28.65 & 4.64 & 3.0 & 1.0 & 90 & $^{+0.05}_{-29.62}$ & 25.9 & $^{+0.69}_{-1.85}$ & 0.76 & 0.01 & 4812$^{+72}_{-72}$	& $\cdots$ & 19 \\
K2-329 & K2 Campaign 12 & 25.31 & 3.73 & 1.9 & 0.5 & 90 & $^{+0.09}_{-28.45}$ & 26.62 & 0.46 & 0.822 & 0.02 & 5282$^{+40}_{-39}$& 1.8$^{+2.2}_{-1.3}$ & 20,21 \\
TOI-1227 & \textit{TESS} & 1.66 & 0.03 & 16.65 & 0.24 & 77.35 & $^{+11.21}_{-5.4}$ & 34.01 & $^{+0.97}_{-1.00}$ & 0.56 & 0.03 & 3072$^{+74}_{-74}$ & 0.01$^{+0.002}_{-0.002}$ & 22 \\
TOI-4562 & \textit{TESS} & 3.81 & 0.03 & 16.4 & 0.3 & 90 & $^{+0.09}_{-7.92}$ & 147.4 & $^{+1.44}_{-1.26}$ & 1.152 & 0.046 & 6096$^{+32}_{-32}$ & 0.3$^{+1}_{-1}$ & 23 \\
V1298 Tau & \textit{TESS}  & 2.97 & 0.06 & 24.8 & 0.2 & 90  & $^{+0.05}_{-8.55}$  & 27 & 1.1 & 1.34 & 0.06 & 4970$^{+120}_{-120}$ & 0.02$^{+0.004}_{-0.004}$ & 24 \\
K2-290 & $\cdots$ & 6.63 & 0.66 & 6.9 & $^{+0.5}_{-0.6}$ & 37.1 & $^{+8.24}_{-6.26}$ & 43.5 & 1.2  & 1.51 & 0.08 & 6302$^{+120}_{-120}$  & 4$^{+1.6}_{-0.8}$& 27,28 \\
WASP-84 & $\cdots$ & 14.36 & 0.35 & 2.56 & 0.08 & 70.82 & $^{+13.37}_{-7.83}$ & 21.70 & 0.72 & 0.77 & 0.02 & 5280$^{+80}_{-80}$& 2.1$^{+1.6}_{-1.6}$ & 25,26 \\
\hline
\multicolumn{15}{c}{Hot Jupiters ($a$/$R_*$ $<$ 20)} \\
\hline
CoRoT-2 & $\cdots$  & 4.52  & 0.02 & 10.3 & 0.9 & 90 & $^{+0.09}_{-17.2}$ & 18.92 & $^{+2.15}_{-2.42}$& 0.9 & 0.02 & 5598$^{+50}_{-50}$& 2.66$^{+1.62}_{-1.62}$ & 29,30 \\
CoRoT-18 & $\cdots$ & 5.53 & 0.33 & 8 & 1 & 90 & $^{+0.05}_{-22.65}$ & 7.01 & $^{+0.28}_{-0.38}$ & 0.88 & 0.03 & 5440$^{+100}_{-100}$& 10.69$^{+3.82}_{-3.82}$& 31 \\
EPIC 246851721 & $\cdots$ & 1.14 & 0.06 & 74.92 &$^{+0.62}_{-0.60}$ & 90 & $^{+0.05}_{-11.48}$& 9.59 & 0.23 & 1.62 & 0.04 & 6202$^{+50}_{-52}$ & 3.02$^{+0.44}_{-0.46}$ & 32 \\
HAT-P-20 & $\cdots$ & 14.48 & 0.02 & 1.85 & 0.27 & 52.63 & $^{+16.43}_{-11.66}$ & 11.36 & 0.14 & 0.68 & 0.01 & 4595$^{+45}_{-45}$ & 6.7$^{+5.7}_{-3.8}$ & 33 \\
HAT-P-22 & $\cdots$  & 28.7 & 0.4  & 1.65 & 0.26 & 64.79 & $^{+20.39}_{-10.63}$ & 8.45 & 0.4 & 1.06 & 0.05 & 5314$^{+50}_{-50}$	& 12.4$^{+2.4}_{-2.4}$ & 34 \\
HAT-P-36 & $\cdots$  & 15.3 & 0.4 & 3.12 & 0.75 & 73.16 & $^{+16.75}_{-14.9}$& 4.93 & 0.1 & 1.04 & 0.02 &5620$^{+40}_{-40}$& 6.6$^{+2.9}_{-1.8}$ & 35 \\
HATS-2 & $\cdots$  & 24.98 & 0.04 & 1.5 & 0.5 & 65.06 & $^{+24.08}_{-13.19}$ & 5.51 & 0.14 & 0.9 & 0.02 & 5227$^{+95}_{-95}$ & 9.7$^{+2.9}_{-2.9}$ & 29,36 \\
HD 189733 & $\cdots$ & 11.95 & 0.02 & 3.25 & 0.02  & 90 & $^{+0.05}_{-14.18}$ & 8.98 & 0.33 & 0.75 & 0.03 & 5050$^{+50}_{-50}$ & 6.2$^{+3.4}_{-3.4}$ & 37,38 \\
HD 209458 & $\cdots$  & 10.65 & 0.75 & 4.8 & 0.2 & 60.78 & $^{+14.9}_{-8.37}$ & 8.78 & 0.15 & 1.16 & 0.01 & 6117$^{+50}_{-50}$ & 4$^{+1.2}_{-1.2}$ & 39,40 \\
K2-29 & $\cdots$  & 10.76 & 0.22 & 3.7 & 0.5  & 68.66 & $^{+18.01}_{-9.05}$ & 10.54 & 0.14 & 0.86 & 0.01 & 5358$^{+38}_{-38}$ & 2.6$^{+1.2}_{-2.35}$ &  41\\
Kepler-17 & $\cdots$ & 12.09 & 0.24 & 4.7 & 1  & 90 & $^{+0.09}_{-24.36}$ & 5.7 & $^{+0.14}_{-0.41}$ & 0.98 & $^{+0.02}_{-0.05 }$& 5781$^{+85}_{-85}$ & 2.9$^{+1.5}_{-1.6}$  & 42 \\
Qatar-1 & $\cdots$  & 23.7 & 0.12 & 1.7 & 0.3 & 90 &$^{+0.09}_{-26.34 }$ & 6.25 & 0.16  & 0.8 & 0.02 & 4910$^{+100}_{-100}$ & 8.9$^{+3.7}_{-3.7}$ & 43,44 \\
Qatar-2 & $\cdots$  & 18.5 & 1.9 & 2.8 & 0.5 & 90 &$^{+0.09}_{-19.94}$ & 6.53 & 0.1 & 0.7 & 0.01 & 4645$^{+50}_{-50}$ & 9.4$^{+3.2}_{-3.2}$ & 45,46,47 \\
WASP-4 & $\cdots$ & 22.2 & 3.3 & 2.14 &$^{+0.38}_{-0.35}$ & 90 &$^{+0.05}_{-26.83}$ & 5.48 & 0.15 & 0.91 & 0.02 & 5540$^{+55}_{-55}$ & 7$^{+2.9}_{-2.9}$ & 29,48 \\
WASP-5 & $\cdots$  & 16.2 & 0.4 & 3.2 & 0.3 & 71.5 &$^{+17.06}_{-7.48}$ & 5.42 & 0.22 & 1.09 & 0.04 &  5770$^{+65}_{-65}$ & 5.6$^{+2.2}_{-2.2}$ & 29,49\\
WASP-6 & $\cdots$  & 23.8 & 0.15 & 1.6 &$^{+0.27}_{-0.17}$ & 63.08 &$^{+19.63}_{-10.49}$ & 10.3 & 0.4 & 0.86 & 0.03 & 5375$^{+65}_{-65}$ & 11$^{+3}_{-7}$ & 29,50,51 \\
WASP-8 & $\cdots$ & 15.31 & 0.8 & 1.9 & 0.05 & 35.97 &$^{+4.87}_{-3.73}$ & 18 & 0.43 & 0.98 & 0.02 & 5690$^{+36}_{-36}$ & 4$^{+1}_{-1}$ & 52 \\
WASP-19 & $\cdots$  & 12.13 & 2.1  & 4.4 & 0.9 & 90 &$^{+0.09}_{-28.18}$ & 3.45 & 0.07 & 1.02 & 0.01 & 5460$^{+90}_{-90}$ & 9.95$^{+2.49}_{-2.49}$ & 53,54 \\
WASP-32 & $\cdots$  & 11.6 & 1 & 3.9 &$^{+0.4}_{-0.5}$ & 54.97 &$^{+19.23}_{-11.07}$ & 7.63 & 0.35 & 1.11 & 0.05 & 6100$^{+100}_{-100}$ & 2.22$^{+0.62}_{-0.73}$  & 55,56 \\
WASP-41 & $\cdots$  & 18.41 & 0.05 & 1.6 & 1.1 & 62.22 &$^{+27.37}_{-16.61}$ & 9.95 & 0.18 & 0.89 & 0.01 & 5546$^{+33}_{-33}$ & 9.8$^{+2.3}_{-3.9}$ & 29,57 \\
WASP-43 & $\cdots$  & 15.6 & 0.4 & 2.26 & 0.54 & 90 & $^{+0.09}_{-27.78}$ & 4.92 &$^{+0.09}_{-0.1}$ & 0.67 & 0.01 & 4520$^{+120}_{-120}$ & 7$^{+7}_{-7}$ & 29,33 \\
WASP-52 & $\cdots$ & 17.26 & $^{+0.51}_{-0.39}$ & 2.62 & 0.07 & 90 & $^{+0.05}_{-11.89}$ & 7.23 & 0.21 & 0.79 & 0.02 & 5000$^{+100}_{-100}$ & 10.7$^{+1.9}_{-1.9}$ & 58,59 \\
WASP-69 & $\cdots$ & 23.07 & 0.16 & 2.2 & 0.4 & 90 & $^{+0.09}_{-21.66}$ & 11.97 & 0.44 & 0.81 & 0.03 & 4700$^{+50}_{-50}$ & 7$^{+7}_{-7}$ & 29,60\\
WASP-85 & $\cdots$ & 13.08 & 0.26 & 3.41 & 0.89 & 86.53 & $^{+3.38}_{-27.82}$ & 8.97 & 0.32 & 0.94 & 0.02 & 5685$^{+65}_{-65}$ & 0.5$^{+0.3}_{-0.1}$ & 61\\
WASP-94A & $\cdots$ & 10.48 & 1.6  & 4.2 & 0.5  & 33 &$^{+11.57}_{-7.47}$ & 7.3 &$^{+0.26}_{-0.22}$ & 1.62 &$^{+0.05}_{-0.04}$& 6170$^{+80}_{-80}$ & 2.7$^{+0.6}_{-0.6}$ & 62 \\
Kepler-8 & $\cdots$ & 7.13 & 0.14 & 8.9 & 1 & 57.85 &$^{+14.86}_{-10.58}$ & 6.98 & 0.18 & 1.5 & 0.04 & 6213$^{+150}_{-150}$ & 3.8$^{+1.5}_{-1.5}$ & 53 \\
Kepler-448 & $\cdots$ & 1.29 & 0.03 & 66.43 & $^{+1.00}_{-0.95}$ & 90 & $^{+0.09}_{-15.76}$ & 19.92 & 1.88 & 1.63 & 0.15 & 6820$^{+120}_{-120}$ & 1.4$^{+0.5}_{-0.5}$ & 63\\
WASP-7 & $\cdots$ & 3.68 & 1.23 & 14 & 2  & 44.57 & $^{+29.31}_{-11.16}$ & 9.08 & 0.56 & 1.48 & 0.09 & 6520$^{+70}_{-70}$ & 2.4$^{+1}_{-1}$ & 64 \\
WASP-12 & $\cdots$ & 6.77 & 1.58 & 1.6 & $^{+0.8}_{-0.4}$ & 8.37 & $^{+5.14}_{-3.6}$ & 3.04 & $^{+0.11}_{-0.1}$ & 1.66 &$^{+0.05}_{-0.04}$ & 6313$^{+52}_{-52}$ & 2$^{+0.7}_{-2}$ & 53 \\
WASP-33 & $\cdots$ & 0.52 & 0.05 & 86.63 &$^{+0.32}_{-0.37}$& 36.15 &$^{+4.69}_{-4.09}$ & 3.69 &$^{+0.05}_{-0.1}$ & 1.51 &$^{+0.02}_{-0.03}$ & 7430$^{+100}_{-100}$ & 0.1$^{+0.4}_{-0.09}$ & 65 \\
WASP-62 & $\cdots$ & 6.65 & 0.13 & 9.3 & 0.2  & 72.85 &$^{+12.33}_{-5.99}$ & 9.53 & 0.39  & 1.28 & 0.05 & 6230$^{+80}_{-80}$ & 0.8$^{+0.6}_{-0.6}$ & 66 \\
WASP-76 & $\cdots$ & 9.29 & 1.27 & 1.48 & 0.28 & 9.18 &$^{+2.57}_{-2.11}$ & 4.02 & 0.16  & 1.76 & 0.07 & 6329$^{+65}_{-65}$ & 1.82$^{+0.27}_{-0.27}$ & 67 \\
WASP-121 & $\cdots$ & 3.38 & 0.4 & 13.56 &$^{+0.68}_{-0.69}$ & 39.12 &$^{+8.11}_{-6.3}$ & 3.8 & 0.11 & 1.44 & 0.03 & 6586$^{+59}_{-59}$ & 1.5$^{+1}_{-1}$ & 68 \\
WASP-167 & $\cdots$ & 1.02 & 0.1 & 49.94 & 0.04 & 34.22 &$^{+4.72}_{-3.83}$ & 4.28 & 0.14 & 1.79 & 0.05 & 7043$^{+89}_{-68}$ & 1.54$^{+0.4}_{-0.4}$ & 69 \\
XO-2 & $\cdots$ & 41.6 & 1.1 & 1.07 & 0.09 & 62.36 &$^{+19.4}_{-10.54}$ & 7.79 & $^{+0.36}_{-0.59}$ & 1 & 0.03 & 5332$^{+57}_{-57}$ & 7.8$^{+1.2}_{-1.3}$& 70 \\
XO-6 & $\cdots$ & 1.79 & 0.06 &  48 & 3  & 62.04 & $^{+16.12}_{-9.59}$ & 8.08 & 1.03 & 1.93 & 0.18 & 6720$^{+100}_{-100}$ & 1.88$^{+0.9}_{-0.2}$  & 71 \\
\enddata
\tablecomments{Kepler-447 was observed in \textit{Kepler} Q0-7, Q9-11, Q13-15, Q17. Kepler-486 was observed in \textit{Kepler} Q1-7, Q9-11, Q13-15, Q17. TOI-1227 was observed in \textit{TESS} Sector 11, 12, 38. TOI-4562 was observed in \textit{TESS} Sector 27-39. V1298 Tau was observed in \textit{TESS} Sector 43, 44. The references column incorporates discovery, $R_{*}$, $P_\mathrm{rot}$, Age, and $T_\mathrm{eff}$ references for all systems above. Values not found in the cited references are either taken from TEPCAT (\citealt{Southworth2011}), \citet{2021ApJ...916L...1A}, or \citet{2022PASP..134h2001A}.}
\tablerefs{(1) \citet{Borsato2019}; (2) \citet{Steffen2012}; (3) \citet{Berger2018}; (4) \citet{2012ApJ...750..114F}; (5) \citet{Sanchis-Ojeda2012}; (6) \citet{Masuda2014}; (7) \citet{Libby-Roberts2020}; (8) \citet{Steffen2013}; (9) \citet{2016ApJ...822...86M}; (10) \citet{Sanchis-Ojeda2013}; (11) \citet{Hebrard2013}; (12) \citet{Bonomo2015}; (13) \citet{Schmitt2014}; (14) \citet{Lillo-Box2015}; (15) \citet{Mancini2016}; (16) \citet{2018AJ....155..158B}; (17) \citet{2017MNRAS.464..850G}; (18) \citet{2018MNRAS.475.1765B}; (19) \citet{Livingston2018}; (20) \citet{Rowe2014}; (21) \citet{2021AJ....161...82S}; (22) \citet{2022AJ....163..156M}; (23) \citet{2022arXiv220810854H}; (24) \citet{David2019}; (25) \citet{Anderson2014A}; (26) \citet{Anderson2015A}; (27) \citet{Hjorth2019}; (28) \citet{Hjorth2021PNAS}; (29) \citet{Bonomo2017}; (30) \citet{Torres2012}; (31) \citet{Hebrard2011}; (32) \citet{Yu2018}; (33) \citet{Esposito2017}; (34) \citet{Mancini2018}; (35) \citet{Mancini2015}; (36) \citet{Mohler-Fischer2013}; (37) \citet{HenryWinn2008}; (38) \citet{Cegla2016}; (39) \citet{Maxted2015}; (40) \citet{Santos2020}; (41) \citet{Santerne2016}; (42) \citet{Desert2011}; (43) \citet{Mislis2015}; (44) \citet{Covino2013}; (45) \citet{Dai2017}; (46) \citet{Bryan2012}; (47) \citet{Mocnik2017}; (48) \citet{Sanchis-Ojeda2011Wasp4}; (49) \citet{Triaud2010}; (50) \citet{Gillon2009}; (51) \citet{Tregloan-Reed2015}; (52) \citet{Bourrier2017}; (53) \citet{Albrecht2012}; (54) \citet{Tregloan-Reed2013}; (55) \citet{Brothwell2014}; (56) \citet{Brown2012}; (57) \citet{Southworth2011}; (58) \citet{Rosich2020}; (59) \citet{Chen2020}; (60) \citet{Casasayas-Barris2017}; (61) \citet{Mocinik2016}; (62) \citet{Neveu-VanMalle2014}; (63) \citet{Johnson2017}; (64) \citet{Albrecht2012A}; (65) \citet{Johnson2015}; (66) \citet{Brown2017}; (67) \citet{Ehrenreich2020}; (68) \citet{Bourrier2020}; (69) \citet{Temple2017}; (70) \citet{Damasso2015}; (71) \citet{Crouzet2017}}

\label{tab:full_table}
\end{deluxetable*}
\end{longrotatetable}

\bibliography{sample63}{}
\bibliographystyle{aasjournal}

\appendix
\section{Light Curve Analysis}

Light curve analysis using \emph{Kepler}, K2, \emph{TESS} photometry, along with generalized Lomb-Scargle periodograms and phased light curves.

\begin{figure}[b]
 \hskip -0.8 in
 \gridline{\fig{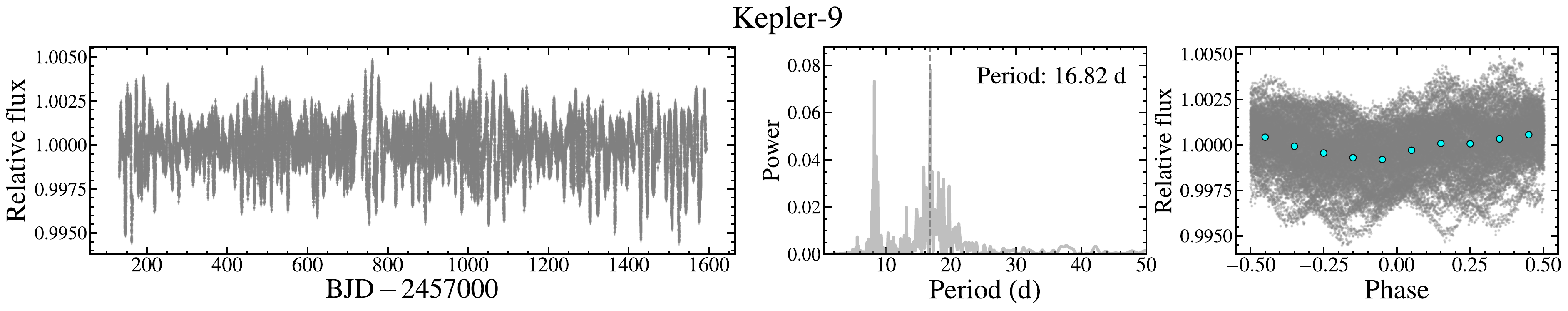}{0.9\textwidth}{}}
 \vskip -.3 in
 \gridline{\fig{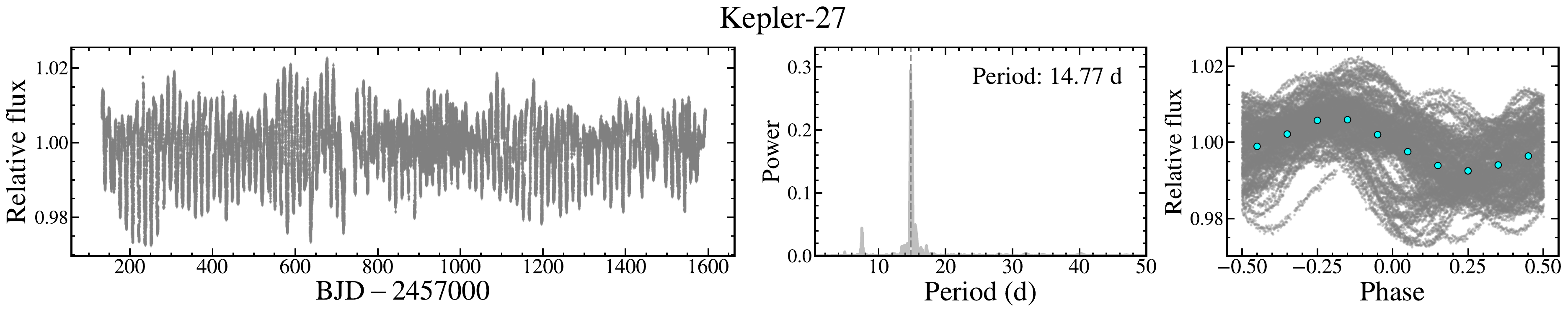}{0.9\textwidth}{}}
 \vskip -.3 in
 \gridline{\fig{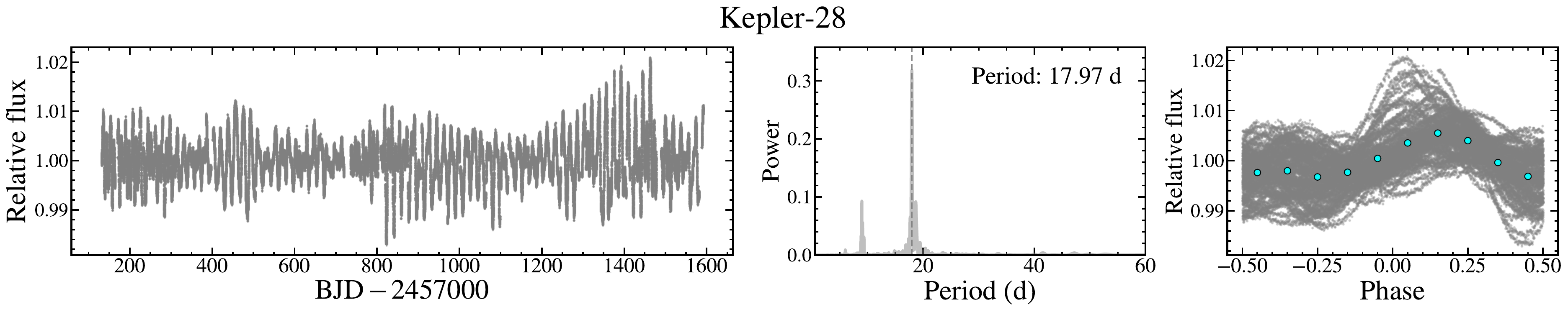}{0.9\textwidth}{}}
 \vskip -.3 in
 \gridline{\fig{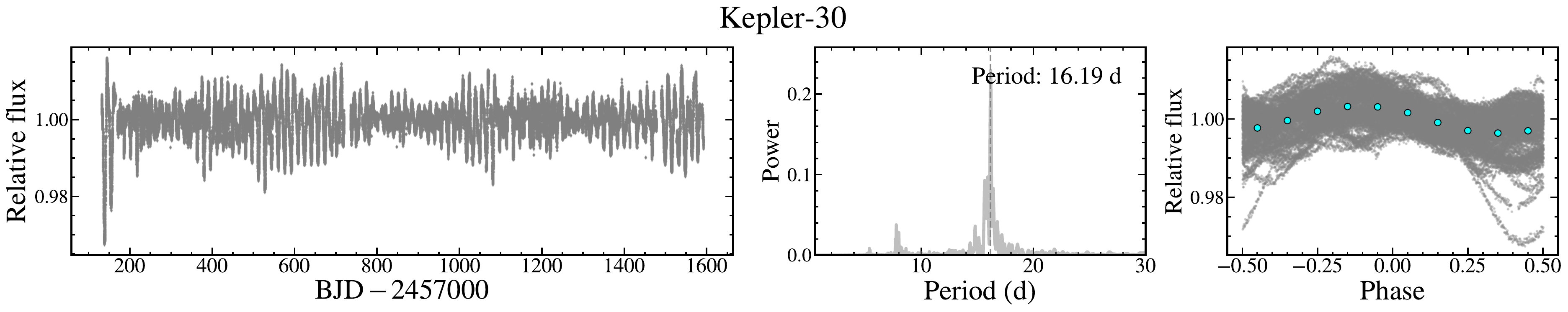}{0.9\textwidth}{}}
 \vskip -.3 in
 \gridline{\fig{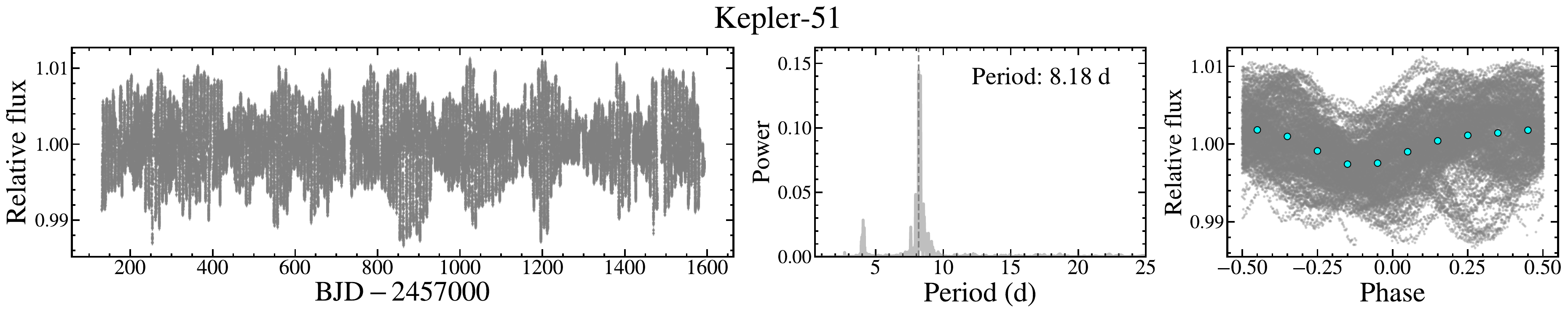}{0.9\textwidth}{}}
 \vskip -.3 in
 \caption{Normalized \emph{Kepler} light curves and GLS periodograms for Kepler-9, Kepler-27, Kepler-28, Kepler-30, and Kepler-51. The strongest periodogram peak is marked with a gray dashed vertical line. The light curve is phased to the period corresponding to the peak and shown in the right panel with the median binned phases displayed in cyan dots.
 \label{fig:lightcurves1}}
\end{figure}

\begin{figure}
 \hskip -0.8 in
 \gridline{\fig{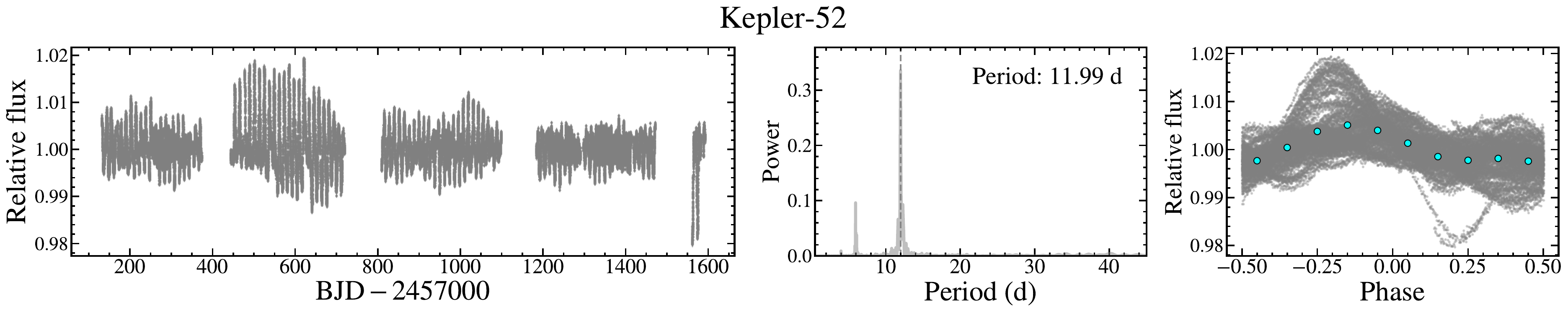}{0.9\textwidth}{}}
 \vskip -.3 in
 \gridline{\fig{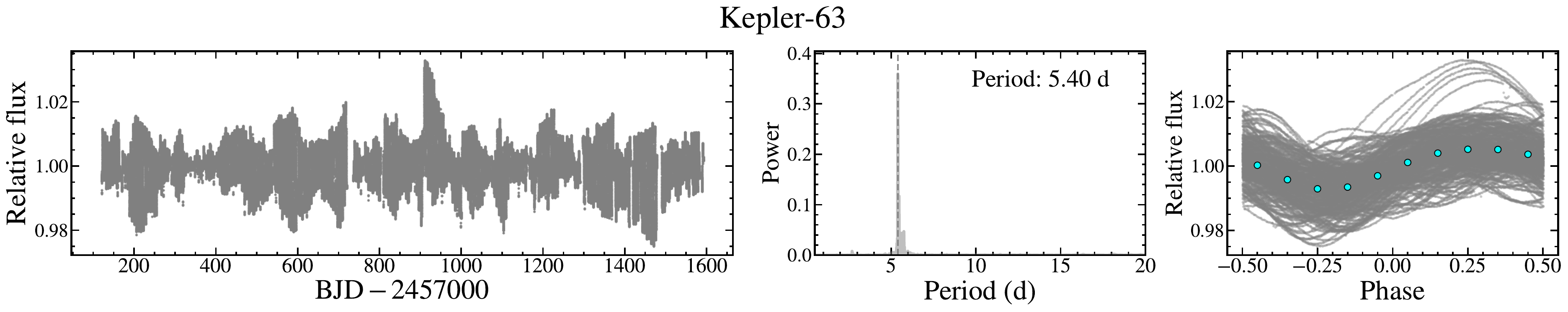}{0.9\textwidth}{}}
 \vskip -.3 in
 \gridline{\fig{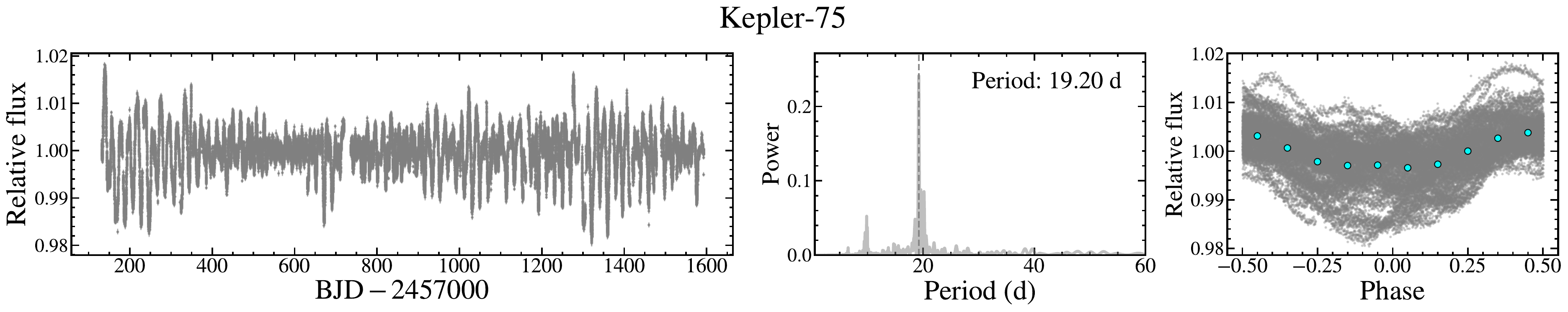}{0.9\textwidth}{}}
 \vskip -.3 in
 \gridline{\fig{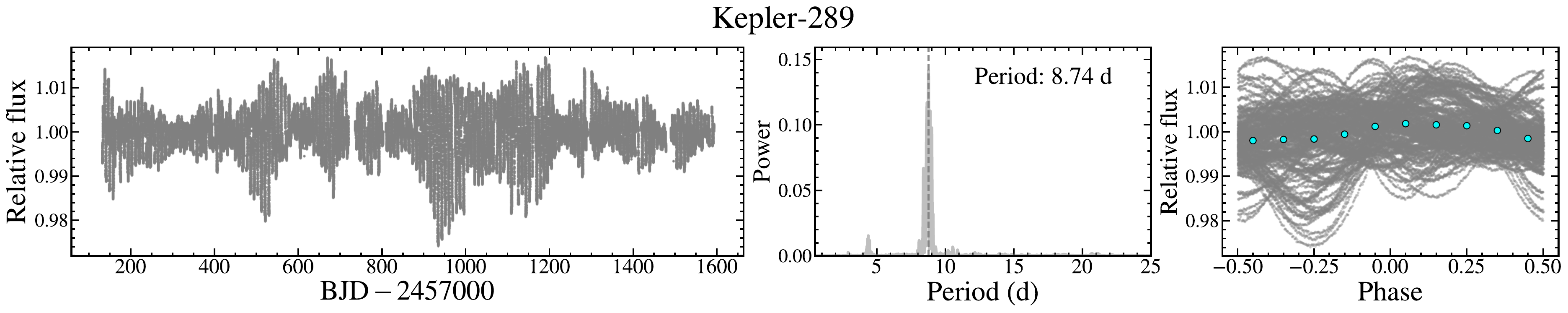}{0.9\textwidth}{}}
 \vskip -.3 in
 \gridline{\fig{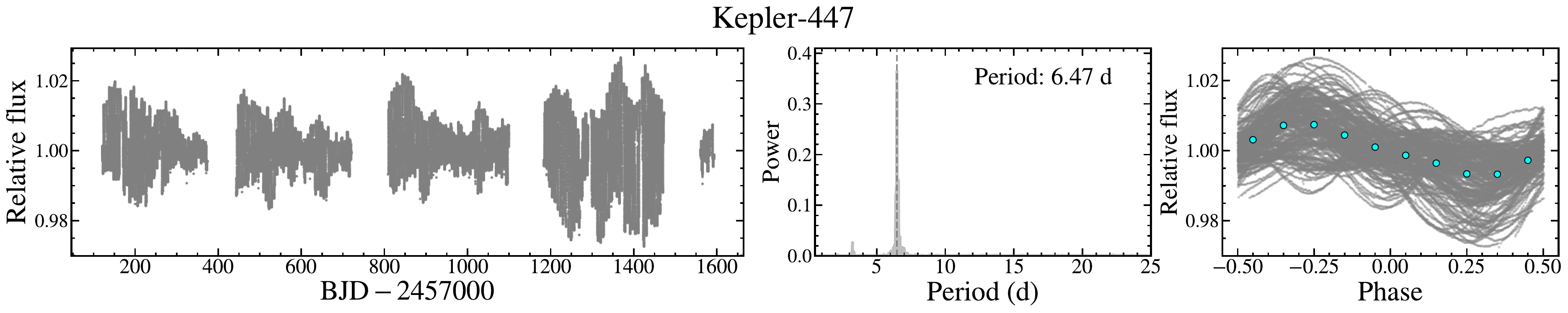}{0.9\textwidth}{}}
 \vskip -.3 in
 \gridline{\fig{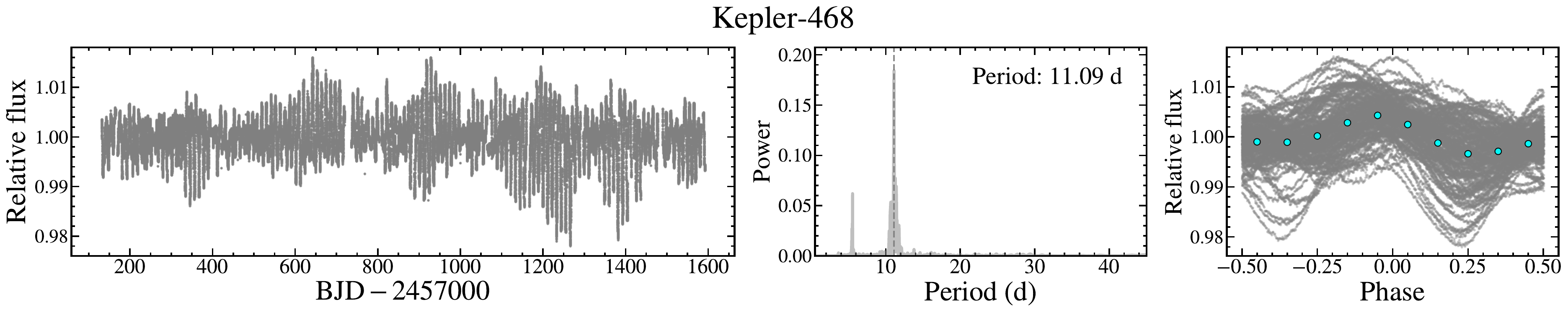}{0.9\textwidth}{}}
 \vskip -.3 in
\caption{Normalized \emph{Kepler} light curves and GLS periodograms for Kepler-52, Kepler-63, Kepler-75, Kepler-289, Kepler-447, and Kepler-468.
 \label{fig:lightcurves2}}
\end{figure}

\begin{figure}
 \hskip -0.8 in
 \gridline{\fig{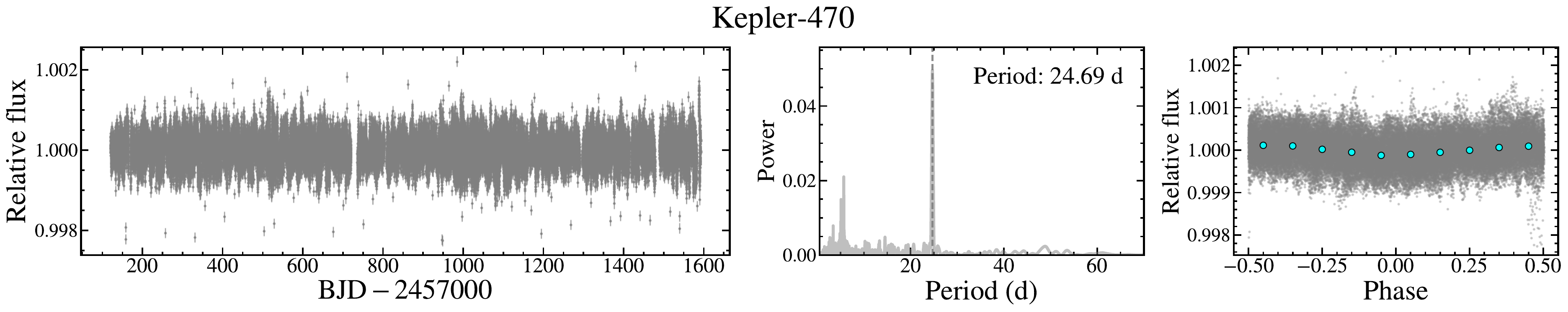}{0.9\textwidth}{}}
 \vskip -.3 in
 \gridline{\fig{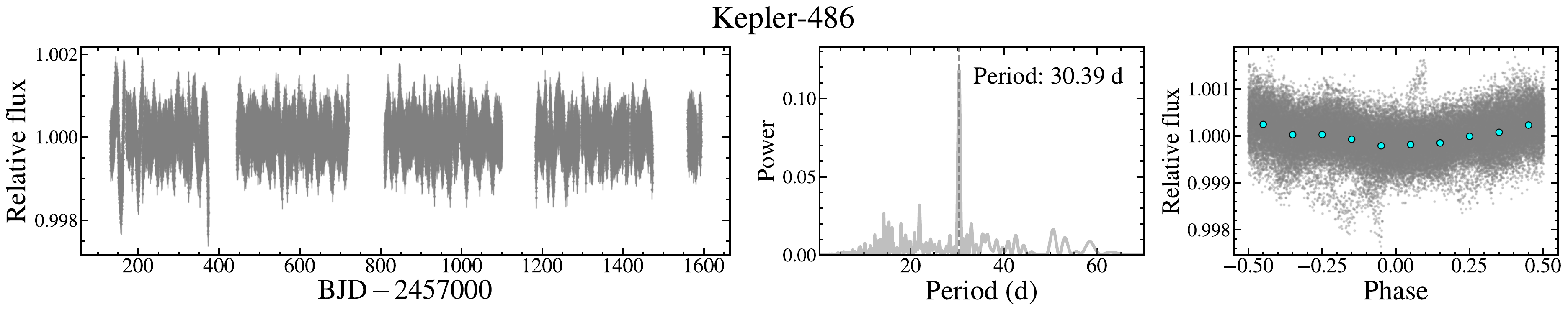}{0.9\textwidth}{}}
 \vskip -.3 in
 \gridline{\fig{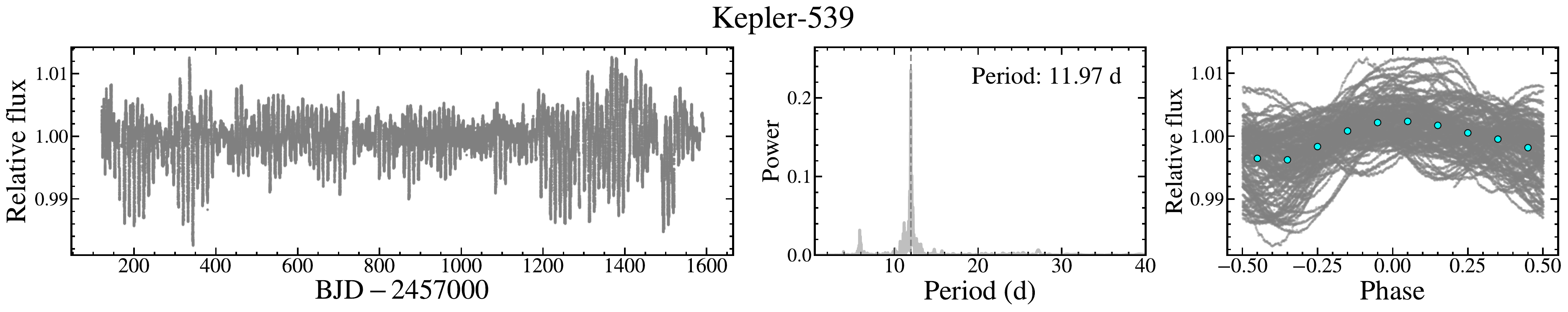}{0.9\textwidth}{}}
 \vskip -.3 in
 \gridline{\fig{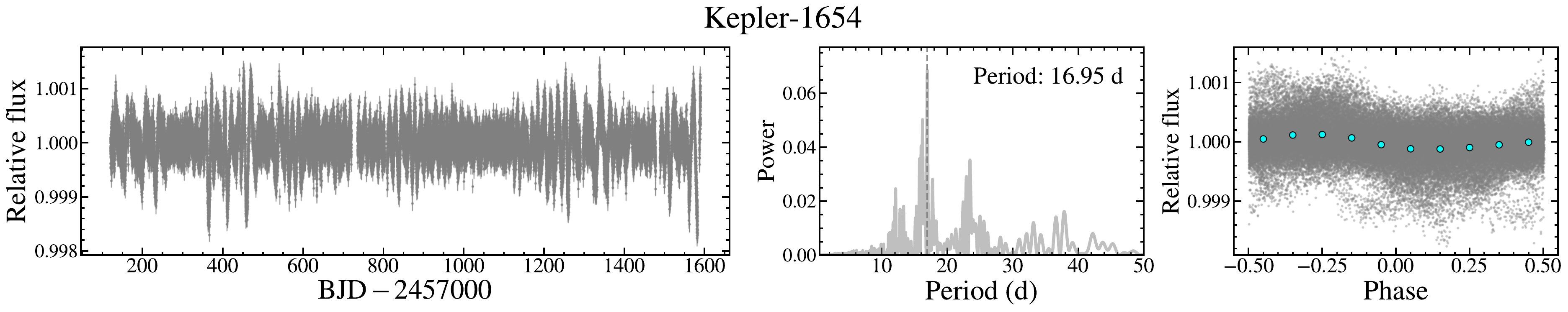}{0.9\textwidth}{}}
 \vskip -.3 in
 \gridline{\fig{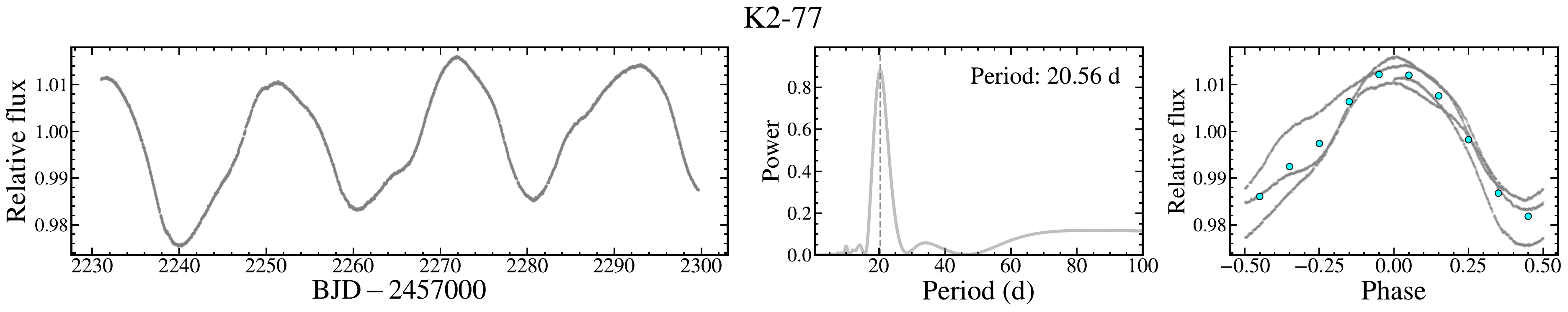}{0.9\textwidth}{}}
 \vskip -.3 in
 \gridline{\fig{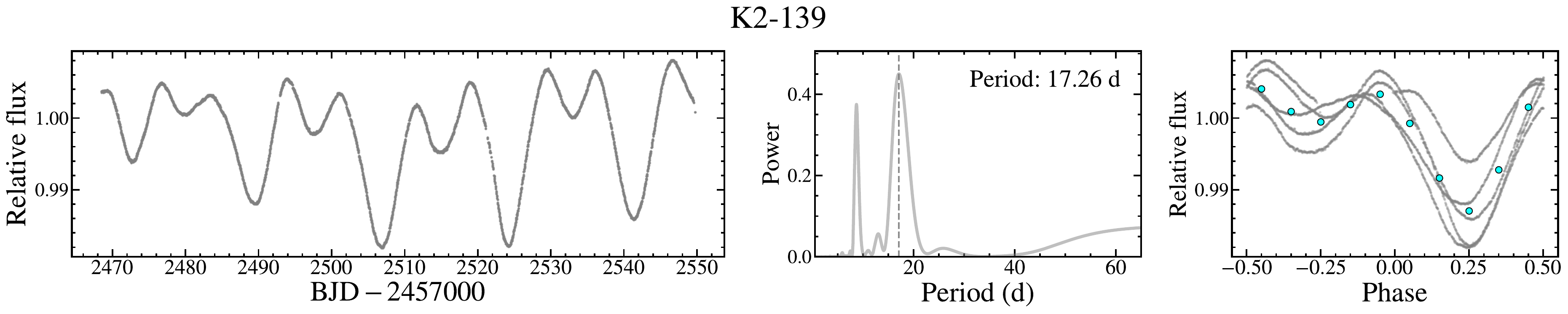}{0.9\textwidth}{}}
 \vskip -.3 in
\caption{Normalized \emph{Kepler} light curves and GLS periodograms for Kepler-470, Kepler-486, Kepler-539, Kepler-1654, K2-77, and K2-139.
 \label{fig:lightcurves3}}
\end{figure}

\begin{figure}
 \hskip -0.8 in
 \gridline{\fig{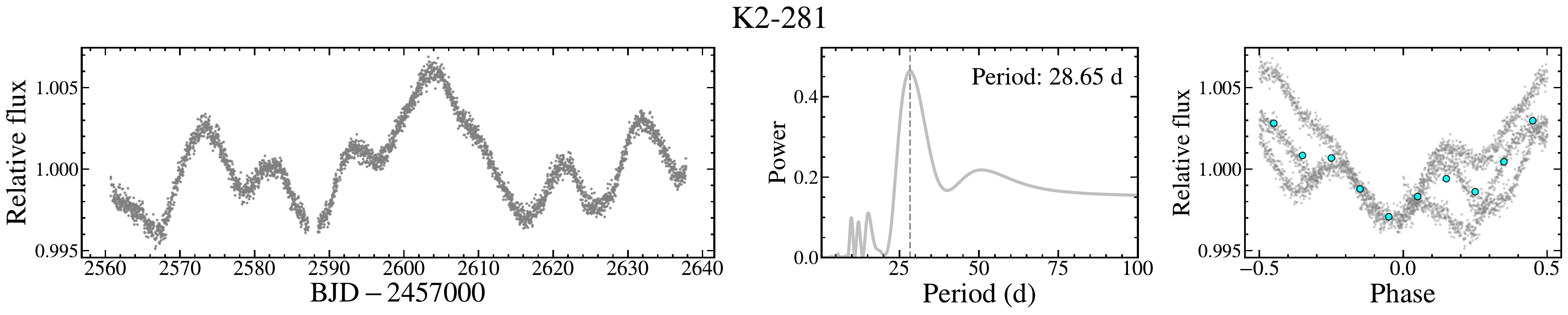}{0.9\textwidth}{}}
 \vskip -.3 in
 \gridline{\fig{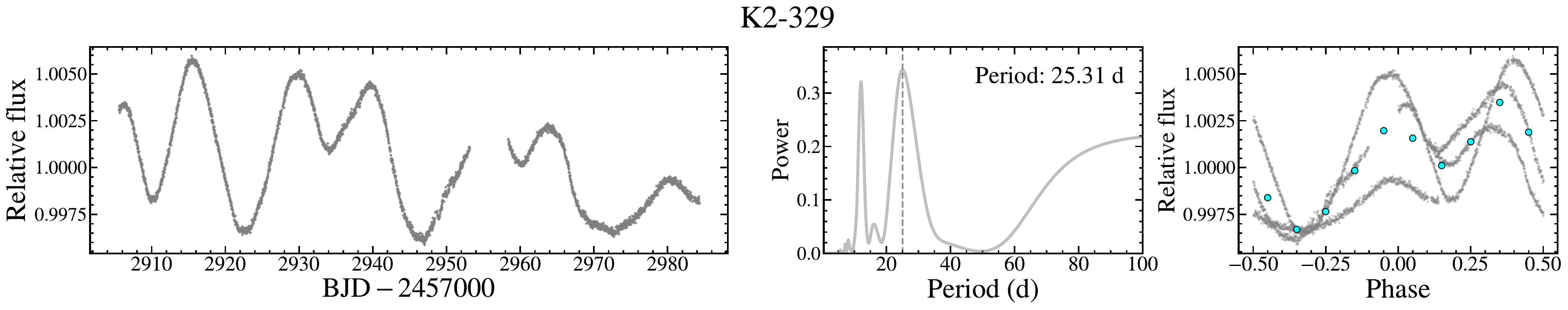}{0.9\textwidth}{}}
 \vskip -.3 in
 \gridline{\fig{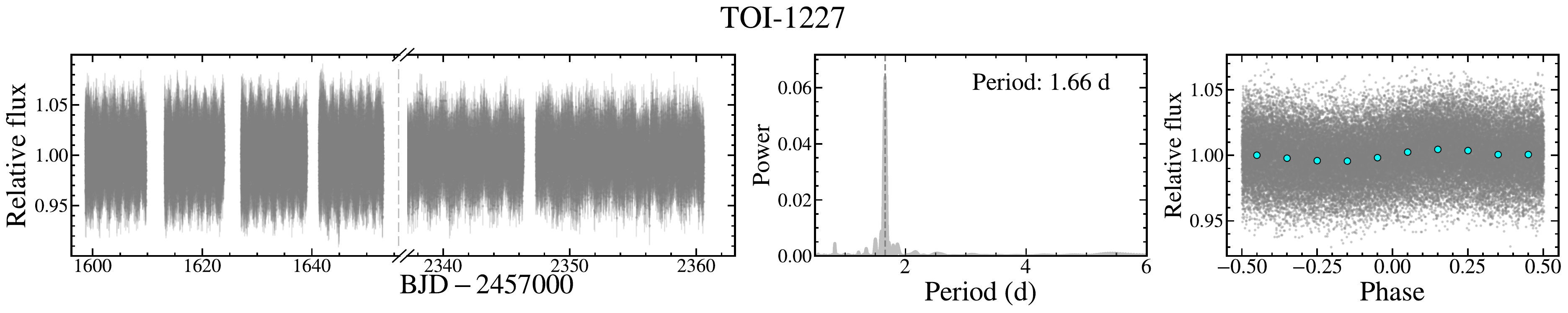}{0.9\textwidth}{}}
 \vskip -.3 in
 \gridline{\fig{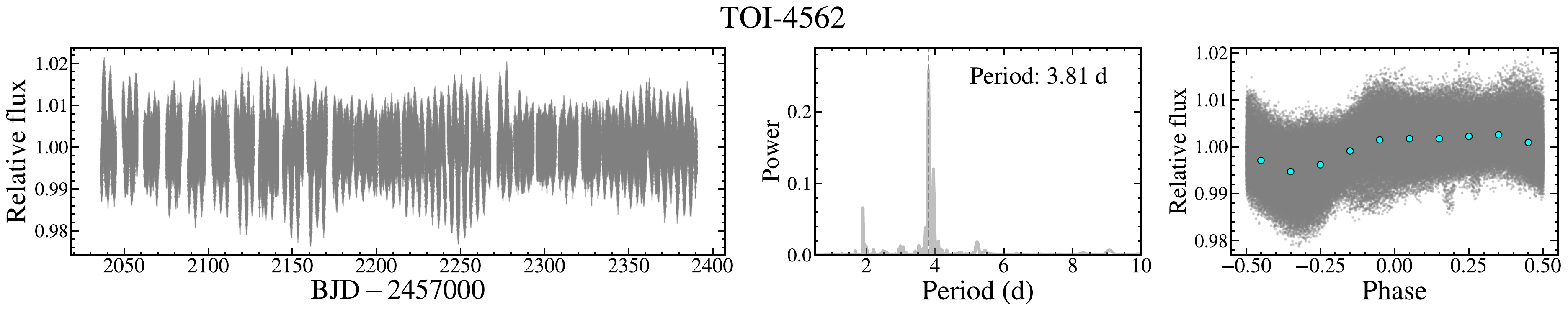}{0.9\textwidth}{}}
 \vskip -.3 in
 \gridline{\fig{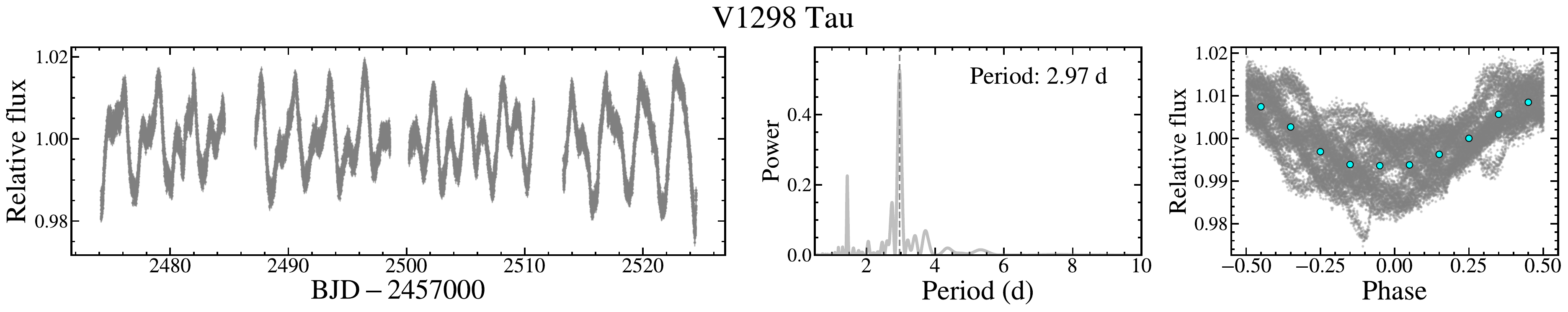}{0.9\textwidth}{}}
 \vskip -.3 in
\caption{Normalized \emph{TESS} and \emph{Kepler} light curves and GLS periodograms for K2-281, K2-329, TOI-1227, TOI-4562, and V1298 Tau.
 \label{fig:lightcurves4}}
\end{figure}

\section{Joint Posterior Distributions} 
\begin{figure}
    \centering
    
    \begin{minipage}[b]{0.48\textwidth}
        \centering
        \includegraphics[width=\linewidth]{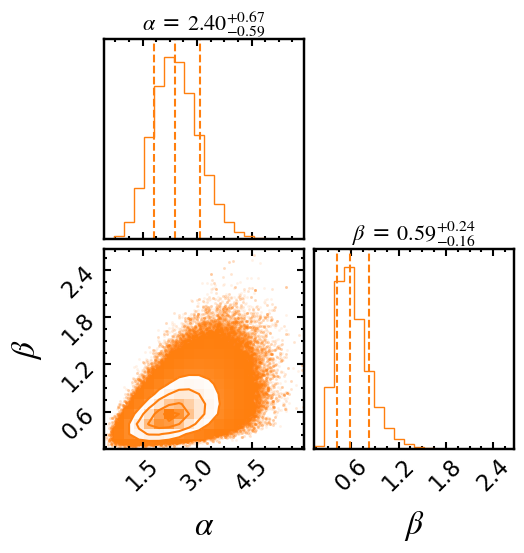}
        \label{fig:plot1}
    \end{minipage}%
    \hfill
    \begin{minipage}[b]{0.48\textwidth}
        \centering
        \includegraphics[width=\linewidth]{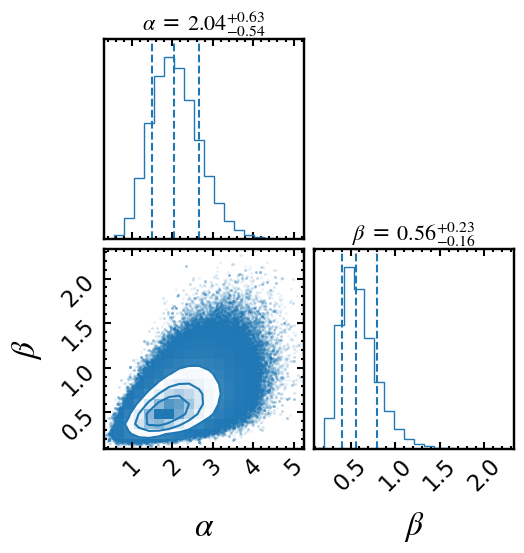}
        \label{fig:plot2}
    \end{minipage}
    \\ 
    \begin{minipage}[b]{0.48\textwidth}
        \centering
        \includegraphics[width=\linewidth]{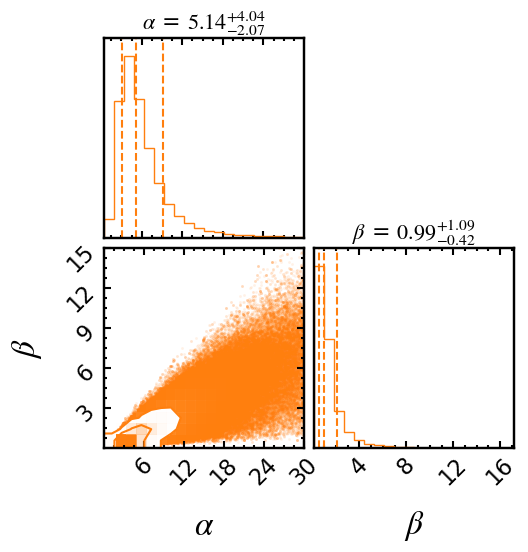}
        \label{fig:plot3}
    \end{minipage}
    \hfill
    \begin{minipage}[b]{0.48\textwidth}
        \centering
        \includegraphics[width=\linewidth]{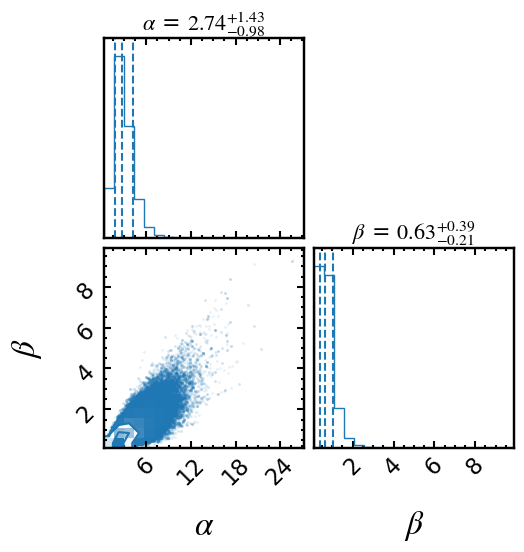}
        \label{fig:plot4}
    \end{minipage}
    
    \caption{Joint posterior distributions of the Beta distribution parameters, $\alpha$ and $\beta$, in our hierarchical Bayesian modeling. Top left: the hot Jupiter sample modeled with a Truncated Gaussian hyperprior. Top right: the warm Jupiter sample modeled with a Truncated Gaussian hyperprior. Bottom left: the hot Jupiter sample modeled with a log-uniform hyperprior. Bottom right: the warm Jupiter sample modeled with a log-uniform hyperprior.}
    \label{fig:four_plots}
\end{figure}

Here we show the joint posterior distributions between the hyperparameters $\alpha$ and $\beta$ of our underlying Beta distribution.  Figure~\ref{fig:four_plots} illustrates how the joint and marginalized distributions are impacted by different hyperpriors between the hot and warm Jupiter host stars.

\end{document}